\documentclass[aps,twocolumn,superscriptaddress]{revtex4-1}

\usepackage{breqn}
\usepackage{graphicx}
\usepackage{dcolumn}   
\usepackage{bm}        
\usepackage{amssymb}   
\usepackage{soul}
\usepackage{gensymb}
\usepackage{enumerate}

\usepackage[breaklinks=true,colorlinks=true,linkcolor=blue,urlcolor=blue,citecolor=blue]{hyperref}

\usepackage{float}

\begin{document}

\title{Global gyrokinetic simulations of ASDEX Upgrade up to the transport time-scale with GENE-Tango}

\author{A.~Di Siena} 
\affiliation{The University of Texas at Austin Austin TX 78712 USA}
\affiliation{Max Planck Institute for Plasma Physics Garching 85748 Germany}
\author{A.~Ba\~n\'on~Navarro}
\affiliation{Max Planck Institute for Plasma Physics Garching 85748 Germany}
\author{T.~Luda}
\affiliation{Max Planck Institute for Plasma Physics Garching 85748 Germany}
\author{G.~Merlo}
\affiliation{The University of Texas at Austin Austin TX 78712 USA}
\author{M.~Bergmann}
\affiliation{Max Planck Institute for Plasma Physics Garching 85748 Germany}
\author{L.~Leppin}
\affiliation{Max Planck Institute for Plasma Physics Garching 85748 Germany}
\author{T.~G\"orler}
\affiliation{Max Planck Institute for Plasma Physics Garching 85748 Germany}
\author{J.~B.~Parker}
\affiliation{Lawrence Livermore National Laboratory Livermore CA 94550 USA}
\author{L.~LoDestro}
\affiliation{Lawrence Livermore National Laboratory Livermore CA 94550 USA}
\author{J.~Hittinger}
\affiliation{Lawrence Livermore National Laboratory Livermore CA 94550 USA}
\author{B.~Dorland}
\affiliation{Department of Physics University of Maryland Maryland MD 20742 USA}
\author{G.~Hammett}
\affiliation{Princeton Plasma Physics Laboratory Princeton NJ 08543 USA}
\author{F.~Jenko} 
\affiliation{Max Planck Institute for Plasma Physics Garching 85748 Germany}
\author{the ASDEX Upgrade Team} 
\affiliation{Max Planck Institute for Plasma Physics Garching 85748 Germany}
\affiliation{See Meyer et al 2019 (https://doi.org/10.1088/1741-4326/ab18b8) for the
ASDEX Upgrade Team.}
\author{the EUROfusion MST1 Team}
\affiliation{See Labit et al 2019 (https://doi.org/10.1088/1741-4326/ab2211) for the
EUROfusion MST1 Team}

\begin{abstract}

An accurate description of turbulence up to the transport time scale is essential for predicting core plasma profiles and enabling reliable calculations for designing advanced scenarios and future devices. This is challenging for high-fidelity gyrokinetic simulations due to the large time-scale separation between microscopic and macroscopic physics in the core of magnetic confinement devices and the already prohibitive computational resources required to cover only the microscopic scales. On the other hand, presently available reduced models fail to correctly describe turbulence in highly electromagnetic regimes and in conditions where supra-thermal particles are substantial. This typically leads to underestimating the on-axis ion temperature for these reduced models, thus affecting the calculations of plasma performance and possibly calling into question predictions of future devices in such conditions. Here, we exploit the gap separation between turbulence and transport time scales and couple the global gyrokinetic code GENE to the transport-solver Tango, including kinetic electrons, collisions, realistic geometries, toroidal rotation and electromagnetic effects for the first time. This approach overcomes gyrokinetic codes' limitations and enables high-fidelity profile calculations in experimentally relevant plasma conditions, significantly reducing the computational cost.

We present numerical results of GENE-Tango for two ASDEX Upgrade discharges, one of which exhibits a pronounced peaking of the ion temperature profile not reproduced by TGLF-ASTRA. We show that GENE-Tango can correctly capture the ion temperature peaking observed in the experiment. By retaining different physical effects in the GENE simulations, e.g., collisions, toroidal rotation and electromagnetic effects, we demonstrate that the ion temperature profile's peaking is due to electromagnetic effects of submarginal MHD instability. Based on these results, the expected GENE-Tango speedup for the ITER standard scenario is larger than two orders of magnitude compared to a single gyrokinetic simulation up to the transport time scale, possibly making first-principles ITER simulations feasible on current computing resources.

\end{abstract}


\maketitle


\section{Introduction}

Reliable calculations of plasma profiles are essential to design scenarios with improved confinement and enable accurate performance predictions in present magnetic confinement devices and future devices. This is a challenging task due to the large variety of coexisting effects acting at different scales in time and space, affecting the evolution of plasma profiles. While particle and energy sources (either externally injected or arising from fusion reactions) act on macroscopic scales, micro-turbulence - driven by plasma instabilities destabilized by gradients in the pressure profile - leads to radial transport of particle and energy on the microscopic scales. A primary goal in fusion research is to develop tools able to describe these mutually interacting effects self-consistently. The main obstacle in this regard is represented by the large timescale separation between microscopic and macroscopic physics in the core of magnetic confinement devices \cite{ITER}. Assuming gyro-Bohm scaling of turbulent transport \cite{Manfredi_PRL_1997}, this timescale gap increases as $(1/\rho_*)^2$, with $\rho_* = \rho_s / a$ being the ratio between the ion sound Larmor radius and the minor radius of the device. Therefore, for a device such as ITER (with $1/\rho_* \sim 1000$) \cite{Shimada_NF_2007}, the microscopic timescales are expected to be six orders of magnitude smaller than the macroscopic timescales at least (while the electron turbulence time scale is expected to be roughly $1 \mu$s, the energy confinement time a few seconds) \cite{Aymar_NF_2001}. This timescale gap has been a major limiting factor for high-fidelity gyrokinetic codes \cite{Brizard_RMP_2007,Garbet_NF_2010} to simulate the micro-turbulence dynamics up to the transport timescale. This is due to the prohibitive computational resources often required by these codes to simulate turbulence, even only at the microscopic timescales. Typically, ion or electron scale gyrokinetic simulations require at least $\sim 10^4 - 10^6$ CPUh at a single radial position for covering only fractions of milliseconds \cite{Mantica_PPCF_2019}. This cost increases substantially (several order of magnitudes) for multi-scale simulations (see e.g., Ref.~\cite{Howard_NF_2015}).

In the past few decades, a major focus has been to develop accurate reduced turbulence models able to grasp the main signatures of plasma micro-instabilities and turbulence at a considerably reduced computational cost. This effort led to the emergence of a hierarchy of different codes mostly based on quasi-linear theory. Among the most well-known and widely used reduced turbulence codes, we note TGLF (trapped-gyro-Landau-fluid) \cite{Staebler_PoP_2007,Staebler_PoP_2016} and QuaLiKiz \cite{Bourdelle_PoP_2007,Bourdelle_PPCF_2015,Citrin_PoP_2012,Citrin_PPCF_2017}. While TGLF is a gyrofluid electromagnetic code retaining shaped tokamak geometry effects, QuaLiKiz is a reduced gyrokinetic code developed in the electrostatic limit and for shifted circular plasmas.

The great advantage of these models is the tremendous speedup (a few seconds on a single CPU) compared to gyrokinetic codes that enabled dynamic profile calculations. This is routinely done nowadays with integrated modelling tools, such as TGLF-ASTRA \cite{Fable_PPCF_2013,Pereverzev_IPP_2002}, TGLF-T-GYRO \cite{Candy_PoP_2009}, QuaLiKiz-JINTRAC \cite{Romanelli_PFR_2014}, by coupling these reduced turbulence models with transport codes. While the turbulence codes compute the transport coefficients for given plasma profiles, the transport codes evolve the profiles according to sources and turbulent fluxes. These tools are shown to accurately reproduce the experimental measurements in different devices and on a large variety of scenarios, contributing to performance predictions and scenario development. 

However, the reduced turbulence models fail to fully describe plasma micro-instabilities and turbulence in strongly nonlinear regimes. In particular, in highly electromagnetic regimes and in cases with substantial external plasma heating, a growing number of studies show that the reduced turbulence codes under-predict turbulence suppression \cite{Doerk_NF_2017,Reisner_NF_2020,Mantica_PPCF_2019,Luda_NF_2021}. This leads to a systematic mismatch of the on-axis temperatures in these regimes, thus affecting plasma performance predictions. This limitation is not expected to affect gyrokinetic codes that are shown to capture correctly nonlinear electromagnetic effects on the microscopic time scales \cite{Holland_NF2012,Citrin_PRL_2013,Whelan_PRL2018,DiSiena_NF_2018,DiSiena_NF_2019}. These findings motivate (i) the further development of the reduced turbulence models and (ii) the improvement of the high-fidelity models aimed at self-consistent profile evolution.

In recent years, an increasing effort has been spent on extending gyrokinetic codes and allowing simulations from the turbulent to the transport timescales. One of the most promising approaches developed so far is the so-called multiple-timescale approach. It employs the basic scheme used with reduced turbulence models and transport codes and consists of simulating turbulence and transport phenomena only on their natural timescales \cite{Sugama_PoP_1997,Sugama_PoP_1998,Abel_2013}. This is achieved e.g., by coupling a gyrokinetic turbulence code to a transport code. More precisely, the turbulence code evaluates the turbulence levels for a given pressure profile over several microscopic or fast-scale characteristic times, while the transport code evaluates the new plasma profiles consistent with the given turbulence levels and the experimental sources. These new profiles are transferred back to the turbulence code, and the process is repeated until the volume-averaged energy and particle injected by the external sources is equal to the turbulent fluxes (steady-state solution). This code coupling leads to a reduction in computational expense of several orders of magnitude, making first-principles simulations over the confinement time – even for ITER-like devices – feasible on presently available computing resources.

The applicability of such a multiple-timescale approach has been already demonstrated when using gyrokinetics directly for the surface-averaged turbulent fluxes in local (flux-tube) frameworks with GENE \cite{Jenko_PoP2000}, GS2 \cite{Kotschenreuther_1995_CPC} and Trinity \cite{Barnes_PoP_2010} and in simplified global setups (circular plasmas and adiabatic electrons) \cite{Parker_NF_2018,Parker_Plasma_2018}. In particular, this global coupling has been achieved by coupling the radially global version of the gyrokinetic code GENE \cite{Goerler_JCP2011} with the transport-solver Tango. In this paper we further extend this new integrated modelling tool (GENE-Tango) by including kinetic electrons, collisions, realistic plasma geometries, electromagnetic effects and toroidal rotation (not yet evolved with Tango). A radially global approach is most likely to be required to investigate global transport features such as internal transport barrier formation \cite{Garbet_PoP_2002,Strugarek_PPCF_2013,DiSiena_PRL_2021}, turbulence spreading \cite{Garbet_NF_1994,Diamond_PoP_1195}, transport avalanches \cite{Pradalier_PRE_2010,McMillan_PoP_2009,Sarazin_PoP_2000}, and supra-thermal particle modes \cite{Chen_RMP_2016}. In particular, GENE-Tango is here applied to study two ASDEX Upgrade discharges. We first benchmark GENE-Tango against previously published results obtained with the radially local model GENE-Trinity \cite{Barnes_PoP_2010} on the ASDEX Upgrade shot $\# 13151$ at $t= 1.35$s. Afterwards, we investigate the shot $\#31555$ at $t= 1.45$s, which exhibits a pronounced peaking of the ion temperature profile that is not captured by TGLF-ASTRA. We show that the electromagnetic GENE-Tango simulations can correctly reproduce the experimental findings. To assess the physical effects responsible for the ion temperature peaking, we perform GENE-Tango simulations retaining different physical effects.

This paper is organized as follows. Section \ref{sec1} briefly introduces the underlying basic equations solved in the gyrokinetic code GENE and transport-solver Tango. The numerical scheme employed to couple these codes (so-called LoDestro method \cite{Crotinger_1997,Shestakov_JCP_2003}) is described in Section \ref{sec2}. While the numerical setup and grid resolutions are discussed in Section \ref{sec3}, the first results with kinetic electrons and collisions with GENE-Tango are presented in Section \ref{sec4} together with a comparison with the profiles computed with ASTRA and GENE-Trinity (local approach). In Section \ref{sec5}, we apply GENE-Tango to study an ASDEX Upgrade H-mode plasma which exhibits a pronounced on-axis peaking not captured by TGLF-ASTRA, showing excellent agreement between the GENE-Tango profiles and the experimental measurements. In Section \ref{impact}, we perform a detailed study of the impact of different physical effects on the evolution of the plasma profiles for this ASDEX Upgrade plasma, focusing in identifying the mechanism responsible for the ion temperature peaking. In particular, the role of collisions (Sections \ref{sec6} and \ref{sec8}), electromagnetic effects (Section \ref{sec9}) and toroidal plasma rotation (Sections \ref{sec7} and \ref{sec9}) - induced by external neutral-beam-injection (NBI) - is analyzed. Section \ref{sec11} contains a comparison of the overall speed-up obtained by running GENE-Tango compared to the macroscopic timescales. Finally, conclusions are drawn in Section \ref{sec12}.

\section{Theoretical framework} \label{sec1}

The high-fidelity model used to study turbulence in magnetic confinement devices for many decades is gyrokinetics \cite{Brizard_RMP_2007,Garbet_NF_2010}. It can be derived rigorously starting from the coupled system of Fokker-Planck and Maxwell's equations under the so-called gyrokinetic ordering \cite{Frieman_PF_1982}. This ordering imposes several constraints - motivated by experimental observations - on the amplitudes and space-time scales of the macro- and micro-physics. In particular, it assumes that the fluctuations are (i) highly anisotropic with parallel correlation length largely exceeding the perpendicular one; (ii) of small amplitude with respect to the background quantities (e.g., magnetic-surface-averaged profiles); and (iii) of low frequency compared to the gyro-frequency with background quantity evolution being significantly slower than the fluctuations. With these assumptions, the particle gyro-motion reduces in the underlying equations by a gyro-ring description. It is therefore particularly convenient to solve the gyrokinetic equations in the so-called guiding-center coordinate system. This leads to a reduction of the dimensions of the distribution function from six to five and the elimination of irrelevant space-time scales with a significant saving in computational time without neglecting any meaningful physical effects. Many gyrokinetic codes, including the gyrokinetic code GENE, exploit the first ordering assumption - particularly valid in the core of magnetic confinement devices - and further reduce the equation complexity by employing the so-called $\delta f$ ordering, namely by splitting the distribution function into a background and a small perturbation $f= f_0 + \delta f_{1}$ (with $\delta f_{1} \ll f_0$). 

Under these assumptions, the system evolves on the microscopic scale as follows 
\begin{dmath}
    \frac{\partial \delta f_{1}}{\partial t} = \frac{q_s}{m_s c}\frac{\partial f_0}{\partial v_\shortparallel}\frac{\partial \bar{A}_{1,\shortparallel}}{\partial t}- \vec{v}_{c} \cdot \left(\vec{\nabla} f_0 - \frac{\mu}{m_s v_\shortparallel} \frac{\partial f_0}{\partial v_\shortparallel} \vec{\nabla}B_0 + \vec{\Gamma}_s \right) - \vec{b}_0 \cdot \left(v_\shortparallel  \vec{\Gamma}_s - \frac{\mu}{m} \vec{\nabla} B_0 \frac{\partial \delta f_{1,s}}{\partial v_\shortparallel} \right) + C\left[\delta f_{1,s}\right].
    \label{eq:eq1}
\end{dmath}
Here, $\vec{\Gamma}_s = \vec{\nabla} \delta f_{1,s} - \left(q_s \vec{\nabla}\bar{\phi}_1/m_s v_\shortparallel\right)\left(\partial f_0/\partial v_\shortparallel \right)$; and $C\left[\delta f_{1,s}\right]$ represents the two-particle Coulomb operator, $\vec{v}_c$ the magnetic drift velocity (containing the curvature, the grad-B and the $\vec{E} \times \vec{B}$ drifts), $B_0$ the toroidal magnetic field, $\vec{b}_0$ the unit vector along the magnetic field, $c$ the speed of light, $\mu = m_s v_\perp^2/(2 B_0)$ the magnetic moment, $q_s$ the charge, $T_{0,s}$ the temperature and $m_s$ the mass of the species s. The overbar on the electrostatic ($\phi_1$) and electromagnetic ($A_{1,\shortparallel}$) fields denotes gyroaveraged quantities. The contribution of the parallel magnetic-field fluctuations is, for the moment, neglected in the global version of the code GENE. We refer the reader to Refs.~\cite{Goerler_JCP2011} for further details.
As mentioned above, this equation is coupled to the Maxwell's equations, which in the particle coordinate system read
\begin{equation}
    -\nabla^2 \phi_1 = 4\pi \sum_s q_s\int \delta f_{1,s}^{pc} d^3v,
    \label{eq:eq2}
\end{equation}
\begin{equation}
    -\nabla^2 A_{1,\shortparallel} = \frac{4 \pi}{c} \sum_s q_s \int v_\shortparallel \delta f_{1,s}^{pc} d^3v.
    \label{eq:eq3}
\end{equation}
Here, $\delta f_{1,s}^{pc}$ represents the perturbed part of the distribution function $f$ in the particle coordinate system. The solution of Eqs.~(\ref{eq:eq1}),~(\ref{eq:eq2}),~(\ref{eq:eq3}) requires the additional field equation to evaluate the time derivative of $ A_{1,\shortparallel}$
\begin{equation}
     -\nabla^2 \frac{\partial A_{1,\shortparallel}}{\partial t} = \frac{4 \pi}{c} \sum_s q_s \int v_\shortparallel \frac{\partial \delta f_{1,s}^{pc}}{\partial t} d^3v.
    \label{eq:eq4}
\end{equation}
This is solved by replacing the time derivative of the perturbed distribution function with the right-hand-side of Eq.~(\ref{eq:eq1}) written in particle coordinates. The transformation between the gyrocenter (removed gyro-angle dependence from guiding center coordinates) and particle variable on the perturbed quantities is performed via the pullback operator. For a detailed description and derivation of the gyrokinetic equations we refer the reader to Ref.~\cite{Brizard_RMP_2007}.

This set of equations is particularly suited to study micro-turbulence dynamics at the microscopic timescales. By solving Eqs.~(\ref{eq:eq1}),~(\ref{eq:eq2}),~(\ref{eq:eq3}),~(\ref{eq:eq4}) we evolve the perturbed distribution function $\delta f_{1}$ on the micro-turbulence time scale and compute particle and energy losses due to plasma micro-instabilities. These are determined by energy and particles fluxes that are respectively defined for the generic species $s$
\begin{equation}
    \overline{Q}_s = \langle {\bf Q_s} \cdot \nabla x \rangle_{S} = \langle \int \frac{1}{2}m_sv^2 \delta f_{1,s} \left({\bf v_{E \times B}} \cdot {\bf \nabla} x \right) d^3v \rangle_{S},
    \label{eq:eq5}
\end{equation}
and
\begin{equation}
    \overline{\Gamma}_s = \langle {\bf \Gamma_s} \cdot \nabla x \rangle_{S} = \langle \int \delta f_{1,s} \left({\bf v_{E \times B}} \cdot {\bf \nabla} x \right) d^3v \rangle_{S}.
    \label{eq:eq6}
\end{equation}
Here, $m_s$ represents the mass of the species $s$, while ${\bf v_{E \times B}}$ the ${\bf E \times B}$ velocity and $\langle \cdot \rangle_{S}$ the surface average.

The background quantities are considered constant on the micro-turbulence time scale since their evolution occurs on time scales that are orders of magnitude slower than the micro-turbulence scales. However, to aim for a self-consistent description of experimental discharges, it is essential to simulate not only the microscopic scales (a few milliseconds in ITER) but the macroscopic scales as well (a few seconds in ITER). This is particularly important since particle and energy losses will inevitably affect plasma profiles which, in turn, will modify the turbulence drive. In this regard, single gyrokinetic simulations covering these long time scales (at least a confinement time) are particularly rare in literature due to the prohibitive computational resources required. Most of the existing work has been carried out on reduced setups (adiabatic electrons, collisionless and/or electrostatic plasma, simplified geometry, no impurities nor supra-thermal particles) \cite{Sarazin_NF_2010,Idomura_PoP_2014,Chang_NF_2017,Wang_NF_2020}. In recent years, an increasing effort has been spent on extending gyrokinetic codes improving code performances, e.g., via GPU porting \cite{Germaschewski_PoP_2021}; and developing new algorithms exploiting the time separation between micro- and macro- physics, e.g., time-telescoping methods \cite{Barnes_PoP_2010,Parker_NF_2018,Sturdevant_PoP_2020}. One of the most promising approaches developed so far allowing gyrokinetic simulations up to the transport time-scale, consists of coupling the gyrokinetic code to a 1D transport code (multiple-timescale approach). In this manuscript, this is done by coupling the global version of the gyrokinetic code GENE to the transport code Tango \cite{Parker_NF_2018,Parker_Plasma_2018}. More precisely, GENE evaluates the turbulence levels for a given pressure profile over several microscopic time steps, while Tango evaluates the new plasma profiles consistent with the given turbulence levels and the experimental sources. For kinetic-electron-setups, the transport code solves the 1D equations for density and pressure (temperature) for each plasma species
\begin{equation}
    \frac{\partial n_s}{\partial t} + \frac{1}{V^'} \frac{\partial}{\partial x} \left(V^' \overline{\Gamma}_s \right) = S_n,
    \label{eq:eq7}
\end{equation}
\begin{equation}
    \frac{3}{2}\frac{\partial p_s}{\partial t} + \frac{1}{V^'} \frac{\partial}{\partial x} \left(V^' \overline{Q}_s \right) = S_{i,e} + \frac{3}{2}n_s \sum_{u=i,e\neq s} \nu_{e,i} \left(T_u - T_s \right),
    \label{eq:eq8}
\end{equation}
with $V^' = d V / d x$ differential volume of the flux surfaces, $\overline{Q}_s$ and $\overline{\Gamma}_s$, respectively, heat and particle fluxes defined in Eqs.~(\ref{eq:eq5}),~(\ref{eq:eq6}), $S_n$ and $S_{i,e}$ external particle and heat sources and $\nu_{e,i}$ collisional energy exchange frequency \cite{Huba2013} defined as
\begin{equation}
    \nu_{e,i} = \frac{3.2 \times 10^{-15} n_e Z_i^2 \Lambda_{e,i}}{m_i T_e^{3/2}}.
    \label{eq:eq9}
\end{equation}
Here, $\Lambda_{e,i} = 10$, the density is expressed in $m^{-3}$, the temperature in $eV$ and the mass in units of proton mass. The collisional energy exchange $\nu_{e,i}$ couples Eq.~(\ref{eq:eq8}) for the ion and electron pressure.

Once the transport code evolves the plasma density and temperatures at the next transport time step, these new profiles are transferred back to the turbulence code, which computes the new turbulence levels consistent with the new profiles. This iterative process is repeated until the volume-averaged energy and particle injected by the external sources is equal to the turbulent fluxes. The multiple time-step approach leads to a significant reduction in computational expense. Interestingly, this approach becomes more and more efficient (with more favorable computational saving) as the time scale separation between turbulence and transport time scales grows. For ITER-like devices, it is expected to speed-up convergence to the steady-state solution by several orders of magnitude, making first-principles simulations over the confinement time feasible on presently available computing resources.

It is worth mentioning that momentum transport and variations to the plasma geometry are not considered in the current version of Tango and neglected for the analyses described within this manuscript. These capabilities will be included in the near future. Moreover, for simplicity, the heat and particle fluxes in Eqs.~(\ref{eq:eq7}),~(\ref{eq:eq8}) consist only of the turbulent contribution, namely Eqs.~(\ref{eq:eq5}),~(\ref{eq:eq6}).

\section{Solution of the transport equations} \label{sec2}

In this Section, we describe the numerical scheme used to couple the global gyrokinetic code GENE and the transport-solver Tango. This coupling is based on the co-called LoDestro method \cite{Crotinger_1997,Shestakov_JCP_2003}. It was first introduced in Ref.~\cite{Crotinger_1997} to evolve the plasma density by coupling the two-dimensional Hasegawa–Wakatani equations to Eq.~(\ref{eq:eq8}). Afterwards, the Hasegawa–Wakatani model was replaced by gyrokinetic simulations in reduced setups involving adiabatic electrons, circular plasmas and simplified shapes for the external sources \cite{Parker_NF_2018,Parker_Plasma_2018}. 

The LoDestro method prescribes a procedure to solve the system of equations \ref{eq:eq7},~\ref{eq:eq8} within an implicit timestep advance. In particular, Eqs.~(\ref{eq:eq7}),~(\ref{eq:eq8}) are discretized in time using backward differences and solved - for each macroscopic timestep - iteratively. The set of Eqs.~(\ref{eq:eq7}),~(\ref{eq:eq8}) is hence written
\begin{equation}
    \frac{n_{m,l} - n_{m-1}}{\Delta t} + \frac{1}{V^'}\frac{\partial}{\partial x} \left(V^' \overline{\Gamma}_{m,l} \right) = S_{m,n},
    \label{eq:eq10}
\end{equation}
\begin{dmath}
    \frac{3}{2}\frac{p_{m,l} - p_{m-1}}{\Delta t} + \frac{1}{V^'}\frac{\partial}{\partial x} \left(V^' \overline{Q}_{m,l} \right) = S_{m,i,e} + \frac{3}{2}n \sum_{u=i,e\neq s} \nu_{e,i} \left(T_u - T \right).
    \label{eq:eq11}
\end{dmath}
Here, the subscript $m$ denotes the macroscopic time index and the subscript $l$ the iteration index within a timestep $\Delta t$. The species subscript has been removed for simplicity. Moreover, the sum over the index $u$ indicates a sum over all the plasma species $s$. It is worth mentioning that in this work, although Eqs.~(\ref{eq:eq10}),~(\ref{eq:eq11}) are fully time-dependent, we solve them only for its steady-state solution taking a single large timestep. As mentioned above, the ion and electron pressure equations are coupled via the collisional energy exchange. Therefore, the ion and electron pressures are updated together implicitly to achieve stable iteration.

The coupled set of Eqs.~(\ref{eq:eq10}),~(\ref{eq:eq11}) are solved in the LoDestro method by setting ad-hoc rules for the turbulent fluxes that depend nonlinearly on the plasma pressure and its logarithmic gradient. In particular, particle and heat fluxes are decomposed in a convective and diffusive contribution, which are assumed to fulfill a local dependence with the pressure and its gradient, namely
\begin{equation}
    \overline{Q}_{m,l}=-D_{m,l-1}^p \partial_x p_{m,l} + c_{m,l-1}^p p_{m,l},
    \label{eq:eq12}
\end{equation}
\begin{equation}
    \overline{\Gamma}_{m,l}=-D_{m,l-1}^n \partial_x n_{m,l} + c_{m,l-1}^n n_{m,l}.
    \label{eq:eq13}
\end{equation}
Here, we have introduced, respectively, the diffusive and convective transport coefficients $D_m$ and $c_m$. They are computed at the iteration $l-1$ and employed to decompose the turbulent fluxes at the iteration $l$.

The splitting of the turbulent fluxes in its convective and diffusive contributions is performed in Tango with a user-specified parameter called $\theta_m$. We have
\begin{equation}
    D_{m,l-1} = - \frac{\theta_{m} \overline{Q}_{m,l-1}}{\partial_x {p}_{m,l-1}},
    \label{eq:eq14}
\end{equation}
and
\begin{equation}
    c_{m,l-1} = \frac{\left( 1- \theta_{m} \right) \overline{Q}_{m,l-1}}{p_{m,l-1}}.
    \label{eq:eq15}
\end{equation}
Given the mostly diffusive nature of turbulent transport in the core of magnetic confinement devices, we have fixed this parameter to $\theta_m = 0.95$ when $D_m > 0$ ($c_m > 0$) and to $\theta_m = -1$ for $D_m < 0$ ($c_m < 0$). We note that the convective and diffusive coefficients $D$ and $c$ should not be interpreted as physical diffusive and pinch terms. At convergence, Eqs.~(\ref{eq:eq12}) -~(\ref{eq:eq15}) provide the turbulent fluxes.

With this splitting of the turbulent fluxes, the discretized transport equations \ref{eq:eq10},~\ref{eq:eq11} can be written for the generic species $s$ as
\begin{dmath}
    \frac{n_{m,l} - n_{m-1}}{\Delta t} + \frac{1}{V^'} \frac{\partial }{\partial x}\left[V^' \left(\theta_{m} \overline{\Gamma}_{m,l-1} \frac{\partial_x n_{m,l}} {\partial_x n_{m,l-1}} + \left(1-\theta_m\right)\overline{\Gamma}_{m,l-1} \frac{n_{m,l}}{n_{m,l-1}} \right)\right] = S_{m,n},
    \label{eq:eq16}
\end{dmath}
\begin{dmath}
    \frac{3}{2}\frac{p_{m,l} - p_{m-1}}{\Delta t} + \frac{1}{V^'} \frac{\partial }{\partial x}\left[V^' \left(\theta_{m}        \overline{Q}_{m,l-1} \frac{\partial_x p_{m,l}} {\partial_x p_{m,l-1}} + \left(1-\theta_m\right)\overline{Q}_{m,l-1} \frac{p_{m,l}}{p_{m,l-1}} \right)\right] = S_{m,n} + \frac{3}{2}n \sum_{u=i,e\neq s} \nu_{e,i} \left(T_u - T \right).
    \label{eq:eq17}
\end{dmath}
Therefore, the plasma density and pressure at the iteration $l$ depend only on known quantities, e.g., transport coefficients computed at the iteration $l-1$, external sources and plasma pressure and density (and their gradients) evaluated at the macroscopic timestep $m-1$. For each timestep $m$, the iterations proceed until the turbulent fluxes - computed with the high-fidelity gyrokinetic code GENE - match the volume-average of the external particle and heat sources.

As already discussed in Ref.~\cite{Shestakov_JCP_2003}, the high sensitivity of turbulent transport on changes in the plasma pressure gradients might lead to stability issues within the iteration loop. This is avoided in Tango by applying relaxation on the fluxes (and hence transport coefficients) and plasma pressure. In particular, we replace the turbulent fluxes with the "relaxed" fluxes, namely
\begin{equation}
    \overline{Q}_{m,l-1}\rightarrow \overline{Q}_{m,l-1}^\alpha = \alpha_q \overline{Q}_{m,l-1} + \left(1 - \alpha_q \right)\overline{Q}_{m,l-2}^\alpha,
    \label{eq:eq18}
\end{equation}
and
\begin{equation}
    \overline{p}_{m,l-1}\rightarrow \overline{p}_{m,l-1}^\alpha = \alpha_p p_{m,l-1} + \left(1 - \alpha_p \right)\overline{p}_{m,l-2}^\alpha.
    \label{eq:eq19}
\end{equation}
This relation procedure acts as an average, and the effective turbulent fluxes seen by the transport code Tango at the iteration $l-1$ correspond to the fluxes computed by the turbulence code GENE at the iteration $l-2$ (with the profiles $p_{l-2}$) and a minor correction that is determined by the fluxes at the present iteration (fixed by the amplitude of $\alpha$). Similarly, the pressure profiles passed by the transport code Tango to the turbulence code GENE at the iteration $l-1$, are evaluated as the average of the pressure profiles at the iteration $l-2$ and the one at the present iteration. The magnitude of the relaxation coefficients depends on the sensitivity of the turbulent transport on the gradients of the plasma pressure and needs to be properly tuned case by case. The optimal values found for the present studies range from $\alpha_p = [0.05 - 0.2]$ and $\alpha_q = [0.1 - 0.4]$.

\section{Numerical setup} \label{sec3}

In the following, we summarize the main parameters employed throughout this work. The GENE simulations are performed with deuterium ions and kinetic electrons with realistic ion-to-electron mass ratio. Collisions are retained (except for the analyses of Section \ref{sec8}) and modelled with a linearized Landau–Boltzmann collision operator with energy and momentum conserving terms \cite{Crandall_CPC_2020}. To account for the effect of impurities - neglected as an active species in this study - we include an effective charge $Z_{\rm{eff}}$. Neoclassical effects are typically small compared to turbulence and are neglected in the turbulent calculations \cite{Connor_PPCF_1993,Physics_NF_2007}. As mentioned previously, the current version of GENE-Tango does not allow a self-consistent evolution of the magnetic geometry as the plasma pressure profiles evolve within the iterations. Therefore, the magnetic equilibrium is fixed to the one reconstructed via CLISTE for each ASDEX Upgrade discharge at the time-slice of interest. 

Different grid sizes and resolutions have been employed in Sections \ref{sec4} and \ref{sec5}. While a radial domain of $\rho_{tor} = [0.1-0.8]$ is used for the analyses of the ASDEX Upgrade discharge $\# 13151$, this is reduced to $\rho_{tor} = [0.05-0.7]$ for the discharge $\# 31555$. Here, $\rho_{tor}$ is the radial coordinate based on the toroidal flux $\Phi$, $\rho_{tor} = \sqrt{\Phi / \Phi_{LCFS}}$. The grid resolution employed in the radial $(x)$, bi-normal $(y)$ and field-aligned $(z)$ directions is respectively $(nx0 \times nky0 \times nz0) = (256\times 48 \times 32$) for the ASDEX Upgrade discharge $\# 13151$ and $(nx0 \times nky0 \times nz0) = (225\times 36 \times 32$) for the discharge $\# 31555$. The discretized toroidal mode number is given by $n = n_{0,min} \cdot j$ with $j$ being integer-valued in the range $j = [0,1,2, ..., nky0-1]$ and $n_{0,min} = 2$. The velocity space grids of species ($s$) are set from $-3.5$ to $3.5$ for $v_\shortparallel/v_{th,s}$ and from $0$ to $12$ for the magnetic moment $\mu B_0 / T_s$. Here, $v_{th,s} = (2T_s/ m_s)^{1/2}$ represents the thermal velocity and $B_0$ the on-axis magnetic field. We employed 32 points along the parallel velocity and 24 along the magnetic moment.

The numerical GENE simulations have been performed running GENE in gradient-driven mode. Therefore, Krook-type particle and heat operators have been applied to keep the plasma profiles (on average) fixed to the ones provided by Tango at each iteration. However, as GENE-Tango approaches the steady-state solution, the amplitude of the Krook heat ($\gamma_k$) and particle ($\gamma_p$) coefficients can be significantly reduced. In particular, we used, respectively, $\gamma_k = 0.01 c_s / a$ and $\gamma_p = 0.01 c_s /a$. Here, $c_s = (T_e / m_i)^{1/2}$ represents the sound speed, with $T_e$ the electron temperature at the reference radial position and $m_i$ the bulk ion mass in proton units.

Numerical fourth-order hyperdiffusion is used to damp fluctuations at large toroidal mode numbers due to electron temperature gradient (ETG) modes to reduce the otherwise prohibitive computational cost of these simulations \cite{Pueschel_CPC_2010}. Moreover, we used buffer regions covering $10\%$ of the GENE simulation radial domain to damp fluctuations to zero near the domain boundaries and enforce consistency with the Dirichlet boundary conditions. In these areas we applied a Krook operator $\gamma_b$ with an amplitude $\gamma_b = 1.0 c_s/a$.

In Tango, different approaches are used to treat the turbulent fluxes and sources in these regions to account for the unphysical damping on plasma fluctuations. In the inner buffer region, the physical sources are set to zero (see e.g., Fig.~\ref{fig:source_13151},~\ref{fig:sources_31555}). This is typically a minor correction to the volume integral of the injected sources due to the reduced plasma volume close to the magnetic axis. On the other hand, at the outer buffer, we introduce extrapolation regions in Tango (which might differ for temperatures and density). Specified radial domains are selected in areas unaffected by the Krook operator. Here, Tango performs an interpolation of the GENE turbulent fluxes. The turbulent fluxes outside this region (towards the right boundary) are replaced by a linear extrapolation until the end of the GENE-Tango grid. The specific choice of the extrapolation regions has been adjusted accordingly during the GENE-Tango iterations. 

Moreover, Tango applies Neumann boundary conditions for the inner boundary and Dirichlet for the outer one.

\section{GENE-Tango benchmark} \label{sec4}

\begin{figure}
\begin{center}
\includegraphics[scale=0.26]{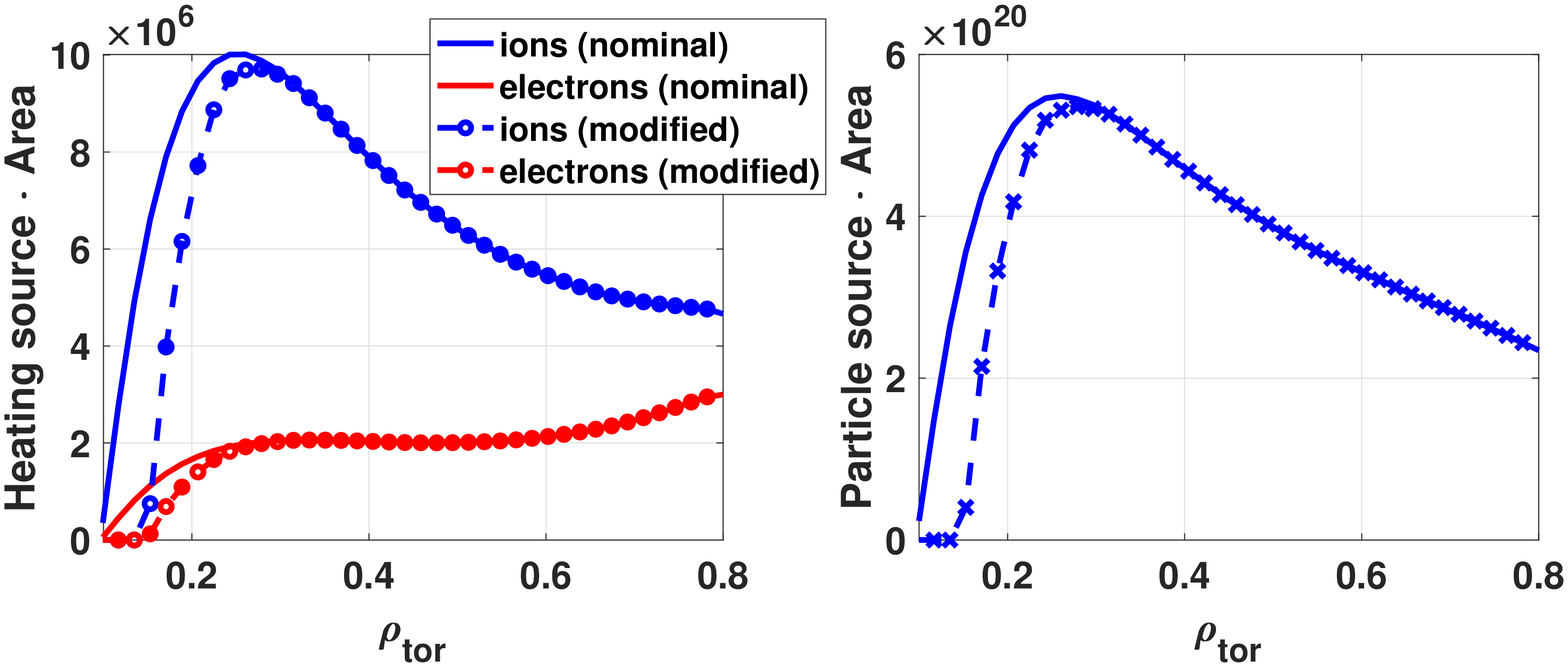}
\par\end{center}
\caption{Radial profiles of the ion and electron heating and particle sources for the ASDEX Upgrade discharge $\# 13151$ at $t = 1.35$s. The continuous lines denote the nominal sources taken from the ITER database \cite{Roach_NF_2008}, while the dotted lines the ones used in GENE-Tango. The amplitude of the sources is set to zero in Tango over the radial domain covered by the inner buffer regions in GENE, i.e., $\rho_{tor} = [0.1 - 0.15]$.}
\label{fig:source_13151}
\end{figure}

\begin{figure*}
\begin{center}
\includegraphics[scale=0.40]{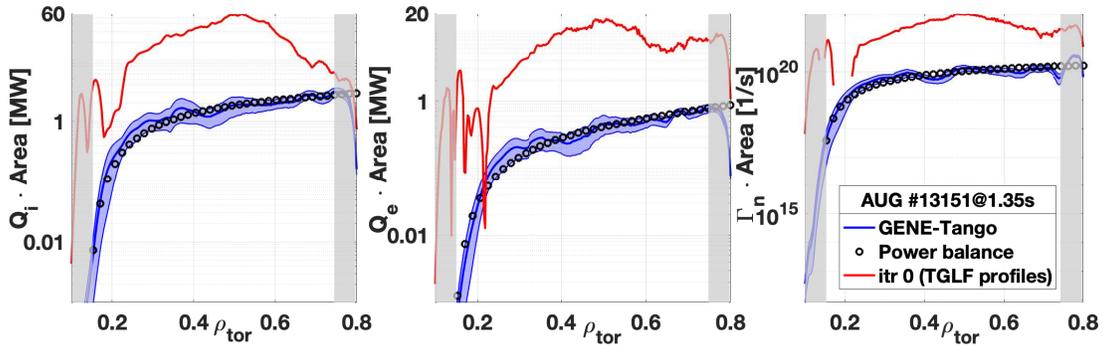}
\par\end{center}
\caption{Time-averaged radial profile of the a) ion, b) electron heat fluxes in MW and c) particle flux in $1/s$ corresponding to the stand-alone GENE simulation with the initial TGLF profiles (red) and the last 5 GENE-Tango iterations (blue). The shaded blue area represents the fluctuations of the turbulent fluxes over the last five iterations. The gray areas denote the buffer regions and the black circles the volume integral of the injected particle and heat sources.}
\label{fig:flux_13151}
\end{figure*}

\begin{figure*}
\begin{center}
\includegraphics[scale=0.40]{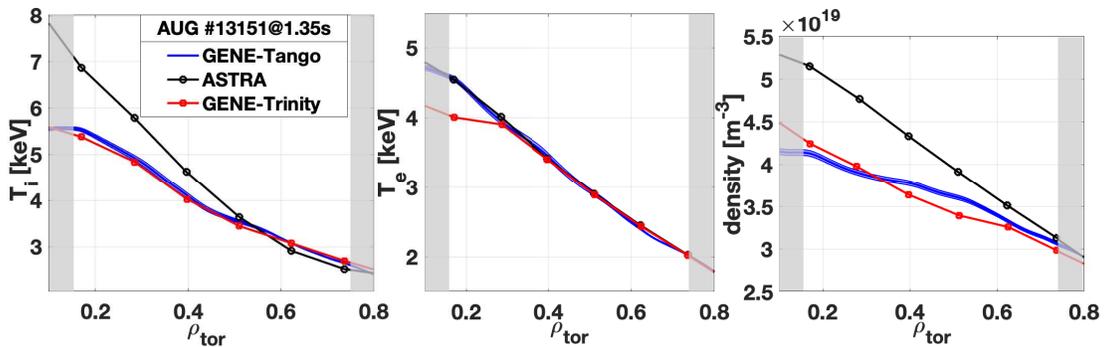}
\par\end{center}
\caption{Comparison of the a) ion, b) electron temperatures and c) density computed by TGLF-ASTRA (black), GENE-Tango averaged over the last five iterations (blue) and GENE-Trinity (red). The shaded blue area represents the fluctuations of the different GENE-Tango profiles over the last five iterations. The gray areas indicate the locations of the buffer regions employed in the GENE simulations.}
\label{fig:pressure_13151}
\end{figure*}

We begin by applying GENE-Tango to study the profile evolution due to external auxiliary power and micro-turbulence for the ASDEX Upgrade discharge $13151$ at $t = 1.35$s \cite{Gruber_NF_2007}. It is an H-mode plasma heated with $5$MW of neutral-beam-injection (NBI). Although a dated experiment, this discharge was selected since it was previously studied with flux-tube simulations with GENE-Trinity \cite{Barnes_PoP_2010}, thus providing an interesting test case to compare different (radially local and global) transport and turbulent models. For the comparison, the same heating, particle source and geometry used in GENE-Trinity are employed in this section. The radial profile of the power absorbed by the thermal ions and electrons and particle refueling are extracted from the ITER database \cite{Roach_NF_2008} and shown in Fig.~\ref{fig:source_13151}. In this regard, the ohmic heating, the total radiated power and the particle source profile from SOL neutrals were not accessible and hence neglected in the simulations. We note that the amplitude of the external sources has been set to zero in the radial domain covered by the inner buffer regions in GENE to avoid nonphysical steepening of the pressure profile in this area. In particular, due to the effect of the Krook operators in GENE the turbulent fluxes will be artificially damped in this region, forcing Tango to increase the pressure gradients. The magnetic geometry is reconstructed by TRACER and read into GENE via numerical field-line tracing provided by the TRACER-EFIT interface \cite{Xanthopoulos_PoP_2009}. Impurities have been neglected for this case and the effective charge is set to $Z_{\rm{eff}}=1$.
The numerical setup and physical parameters are the same as the ones summarized in Section \ref{sec1}. Before starting the coupled GENE-Tango iterations, a GENE standalone simulation is performed to evaluate the turbulence levels corresponding to the initial ion and electron temperatures and density. These profiles are fixed to the ASTRA ones at $t = 1.35$s (see Fig.~\ref{fig:pressure_13151}). This choice sets a constraint over the temperatures and density at the right boundary $\rho_{tor} = 0.8$. The resulting time-averaged (over the steady-state nonlinear phase) heat and particle fluxes are shown in Fig.~\ref{fig:flux_13151}.

With the initial profiles computed by ASTRA, GENE predicts large turbulent fluxes that are not consistent with the volume integral of the injected heating and particle sources. Excellent agreement between turbulence fluxes and volume integral of injected sources is achieved only after running GENE-Tango for $80$ iterations, corresponding to an overall GENE run-time of $41$ms. This is illustrated in Fig.~\ref{fig:flux_13151} where the GENE turbulent heat and particle fluxes - averaged over the last five GENE-Tango iterations - are shown.
To speed up the convergence, the relaxation coefficient $\alpha_p$ and the GENE run-time per iteration have been initially fixed to $\alpha_p = 0.1$ and $t = 250 c_s/a$ and then progressively reduced to $\alpha_p = 0.05$ and $t = 150 c_s /a$ as the GENE-Tango iterations were approaching the final solution. This reduction of $\alpha_p$ is required to make the GENE-Tango iterations more stable, as the system is brought closer to marginal stability. The relaxation coefficient acting on the turbulent fluxes has been kept constant to the value $\alpha_q = 0.4$. 
Looking at steady-state profiles of Fig.~\ref{fig:pressure_13151}, we observe excellent agreement between GENE-Tango and the local GENE-Trinity (flux-tube) model for the ion and electron temperature, with a relative error among the profiles of $\sim 5\%$. The largest deviations are observed on the electron temperature at $\rho_{tor} \approx 0.2$. These differences in the profiles might be caused by finite-size effects, with $130 <1/\rho_* < 210$ in $\rho_{tor} = [0.1-0.8]$ \cite{McMillan_PRL_2010,goerler_PoP_2011}, which get more pronounced as $\rho_{tor}$ is reduced. These essentially global effects typically lead to an over-prediction of the turbulence levels in the flux-tube GENE simulations when compared to the radially global ones \cite{McMillan_PRL_2010,goerler_PoP_2011}, thus causing a more pronounced flattening of the bulk profiles to match the fixed injected power. For a medium-size device as ASDEX Upgrade the flux-tube calculation can lead to an over-prediction of the turbulent fluxes up to a factor of two compared to the global code \cite{Navarro_PoP_2016}. The plasma density also shows also qualitatively good agreement between GENE-Tango and GENE-Trinity, although larger differences ($\sim 7\%$) are observed compared to the temperature profiles. 

\begin{figure*}
\begin{center}
\includegraphics[scale=0.40]{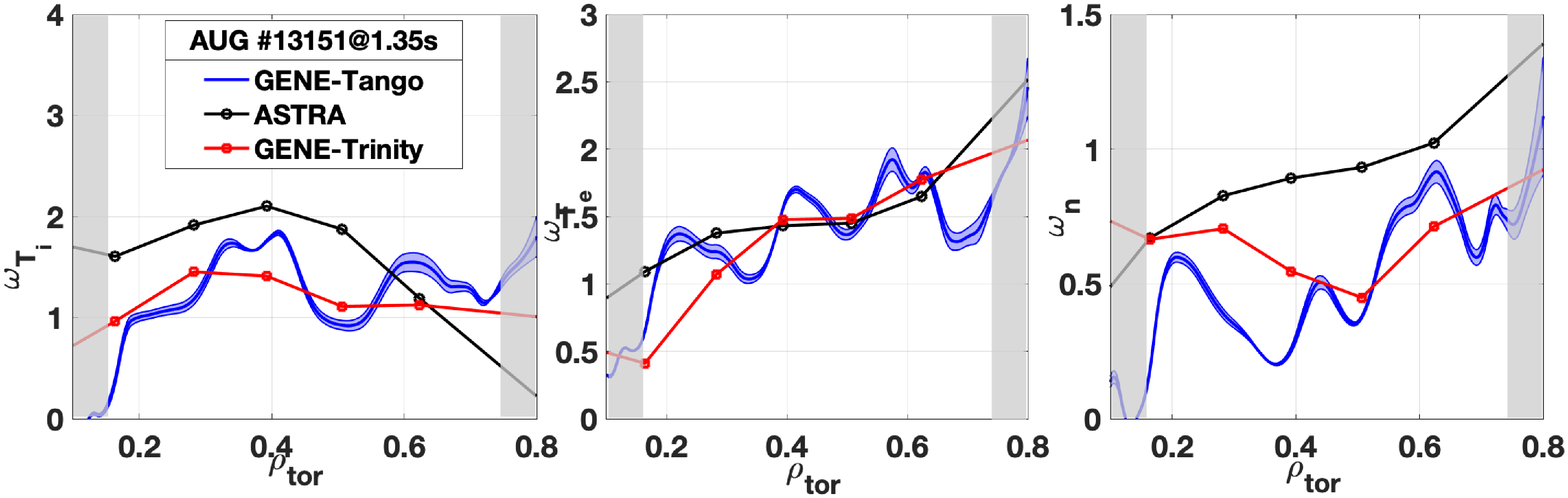}
\par\end{center}
\caption{Comparison of the a) ion, b) electron logarithmic temperatures and c) density gradients computed by TGLF-ASTRA (black), GENE-Tango averaged over the last five iterations (blue) and GENE-Trinity (red). The shaded blue area represents the fluctuations of the different GENE-Tango profiles over the last five iterations. The gray areas indicate the locations of the buffer regions employed in the GENE simulations.}
\label{fig:gradients_13151}
\end{figure*}

Although the GENE-Tango profiles are largely consistent with the GENE-Trinity ones, more pronounced deviations are found with respect to the ion temperature and density computed by ASTRA. In Ref.~\cite{Barnes_PoP_2010} the lack of flow shear in the GENE-Trinity simulations and the oversimplified magnetic equilibrium used in ASTRA (possibly MHD unstable) were proposed as possible causes of discrepancy. On the other hand, the three transport codes' electron temperature profiles are in close agreement, apart from a deviation by Trinity at small $\rho_{tor}$. For completeness we add the comparison of the logarithmic temperature and density gradients in Fig.~\ref{fig:gradients_13151}.

As noted above, these results reveal that the plasma profiles for this ASDEX Upgrade discharge are particularly stiff, characteristic of turbulence near marginal stability. Relatively small differences in the ion temperature and density logarithmic gradients (Fig.~\ref{fig:gradients_13151}) are found to impact the turbulent fluxes considerably, as shown in Fig.~\ref{fig:flux_13151}.

\section{Ion temperature peaking captured with GENE-Tango at ASDEX Upgrade} \label{sec5}

\begin{figure*}
\begin{center}
\includegraphics[scale=0.40]{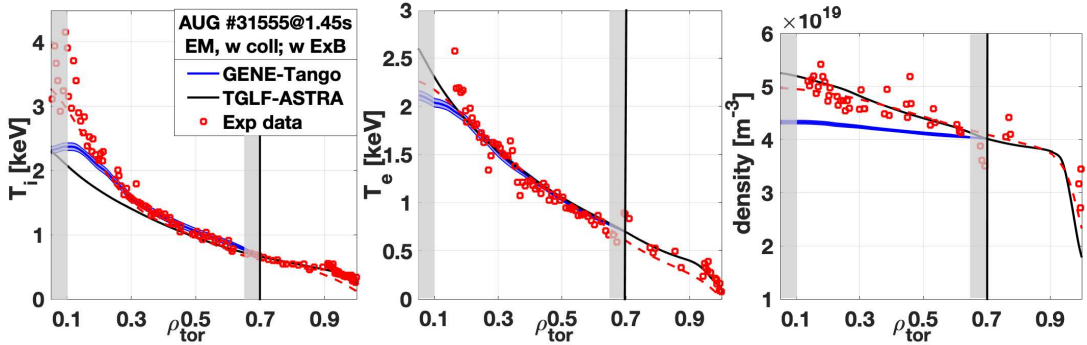}
\par\end{center}
\caption{Comparison of the a) ion, b) electron temperatures and c) density computed by TGLF-ASTRA (black), GENE-Tango averaged over the last five iterations (blue) and experimental measurements (red). The shaded blue area represents the fluctuations of the different GENE-Tango profiles over the last five iterations. The vertical black line delimits the right boundary of the GENE-Tango radial domain. The gray areas indicate the locations of the buffer regions employed in the GENE simulations.}
\label{fig:pressure_em_exb}
\end{figure*}

We apply GENE-Tango to analyze a more recent discharge at ASDEX Upgrade, namely the shot $\# 31555$ at $t = 1.45$s (previously studied in Ref.~\cite{Luda_NF_2021} with TGLF-ASTRA). This experiment exhibits a pronounced peaking of the ion temperature profile not reproduced by the reduced turbulence model TGLF-ASTRA (see, e.g., Fig.~\ref{fig:pressure_em_exb}). This is a known limitation of reduced turbulence models, typically under-estimating the stabilizing effect of electromagnetic effects and supra-thermal particles on ion-scale turbulence, leading to rather flat profiles not consistent with the experimental measurements \cite{Doerk_NF_2017,Reisner_NF_2020,Mantica_PPCF_2019,Luda_NF_2021}.

\begin{figure}
\begin{center}
\includegraphics[scale=0.26]{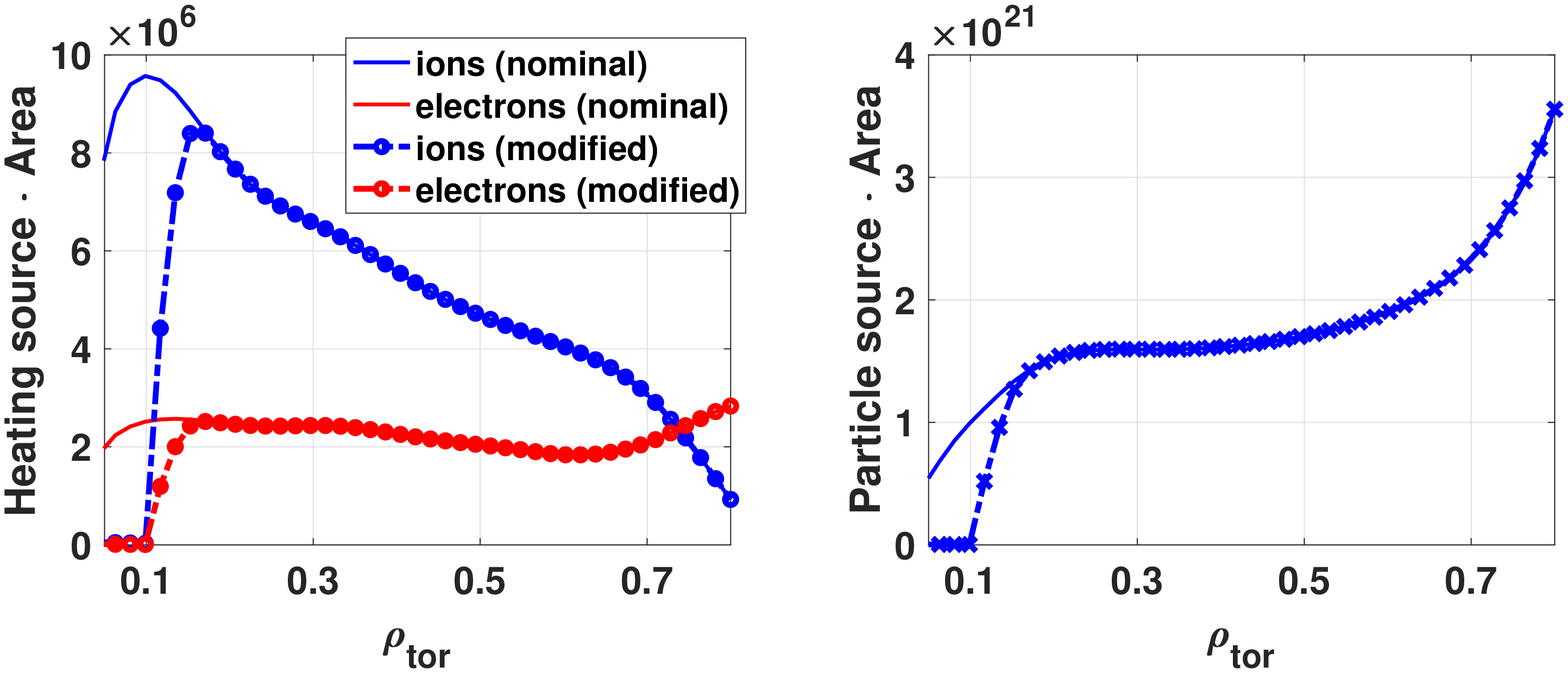}
\par\end{center}
\caption{Radial profiles of the ion and electron heating and particle sources for the ASDEX Upgrade discharge $\# 31555$ at $t = 1.45$s. The continuous lines denote the nominal sources taken from ASTRA, while the dotted lines the ones used in GENE-Tango. The amplitude of the sources is set to zero in Tango over the radial domain covered by the inner buffer regions in GENE, i.e., $\rho_{tor} = [0.05 - 0.10]$.}
\label{fig:sources_31555}
\end{figure}

The ASDEX Upgrade shot $\# 31555$ at $t = 1.45$s  is an H-mode plasma with $5$MW of NBI heating with plasma current $I_p = 0.6$MA, magnetic field $B_t = 2.8$T and relatively low plasma fueling rate $\overline{\Gamma}_n = 0.7 \times 10^{22} e/s$. The power absorbed by the thermal species and particle fueling computed by ASTRA are illustrated in Fig.~\ref{fig:sources_31555}. While the electron heating source is defined as the sum of the ohmic, NBI heating, exchange electrons-ions energy term and the radiated power; the ion heating source as the sum of NBI heating and exchange electrons-ions energy term. The particle source retains the contribution coming from both the NBI source and the SOL neutrals. We note that the amplitude of the external sources has been set to zero in the radial domain covered by the inner buffer regions in GENE to avoid nonphysical steepening of the pressure profile in this area. In particular, due to the effect of the Krook operators in GENE the turbulent fluxes will be artificially damped in this region, forcing Tango to increase the pressure gradients. 

On this specific discharge, we apply GENE-Tango simultaneously retaining electromagnetic effects, collisions and toroidal external rotation in the GENE simulations. The numerical setup and physical parameters are the same as the ones summarized in Section \ref{sec1}.

\begin{figure}
\begin{center}
\includegraphics[scale=0.45]{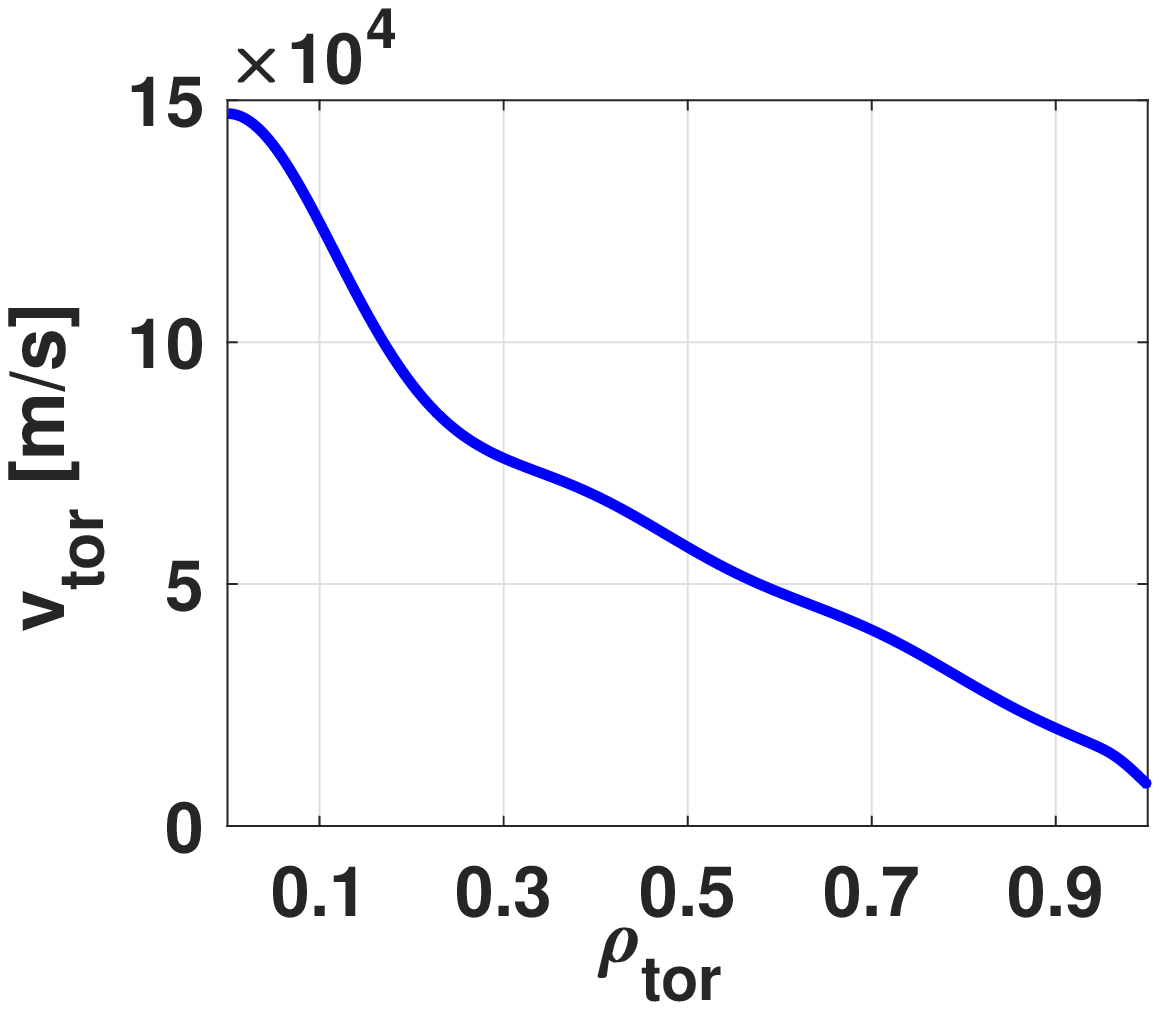}
\par\end{center}
\caption{Toroidal plasma rotation taken from the measurements of the Charge Exchange diagnostic fitted with a third order polynomial for the ASDEX Upgrade discharge $\#31555$ at $t = 1.45$s. Positive values of $v_{tor}$ denote a counter-clockwise rotation.}
\label{fig:vtor}
\end{figure}

As mentioned previously, the current version of GENE-Tango does not evolve the toroidal angular momentum. Therefore, we employed a fixed $v_{tor}$ profile throughout the GENE-Tango iterations. The profile of the toroidal rotation is taken from the measurements of the Charge Exchange diagnostic \cite{Viezzer_RSI_2012}, fitted with a third order polynomial and it is shown in Fig.~\ref{fig:vtor}. Positive values of $v_{tor}$ denote a counter-clockwise rotation.

In the global version of GENE, a phase factor is applied on the perturbed part of the distribution function of each species at every time-step to mimic the effect of finite toroidal rotation, i.e., $f_1 = f_1 {\rm exp}^{-{\rm i} \omega_{E \times B} dt}$. Here, $\omega_{E\times B}$ is defined as follows
\begin{equation}
    \omega_{E\times B} = \frac{v_{tor} \mathcal{C}_y k_y a}{c_s \rho_*},
\end{equation}
with $\mathcal{C}_y = x_0 / q(x_0)$, $x_0 = 0.375$ center of the radial box, and $a$ minor radius of the device.
We note that (global) GENE does not include - at the moment - the parallel flow shear term. Therefore, no model for the parallel velocity gradient (PVG) instability is retained in our simulations. This might possibly lead to an over-estimation of the impact of a finite toroidal rotation on the turbulent fluxes \cite{Kinsey_PoP_2005}.

\begin{figure*}
\begin{center}
\includegraphics[scale=0.40]{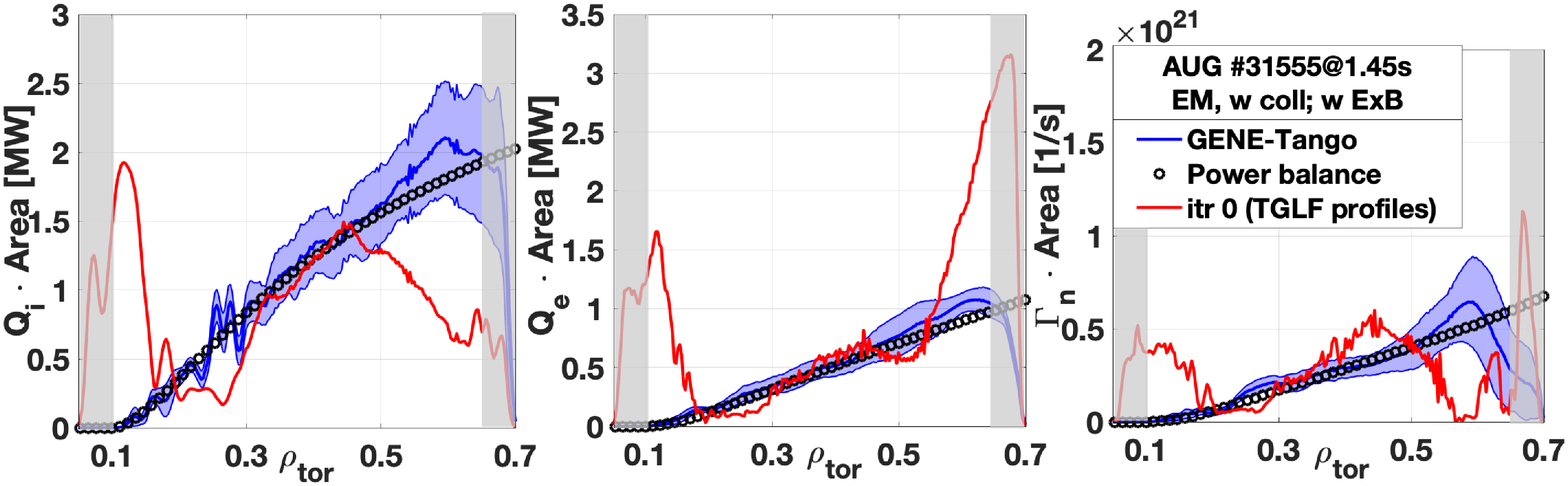}
\par\end{center}
\caption{Time-averaged radial profile of the a) ion, b) electron heat fluxes in MW and c) particle flux in $1/s$ corresponding to the stand-alone GENE simulation with the initial TGLF-ASTRA profiles (red) and the last 5 GENE-Tango iterations (blue). The shaded blue area represents the fluctuations of the turbulent fluxes over the last five iterations. The gray areas denote the buffer regions and the black circles the volume integral of the injected particle and heat sources.}
\label{fig:flux_em_exb}
\end{figure*}

Before starting the Tango iterations, we performed a GENE stand-alone global simulation with the ion and electron temperature and density profiles computed by TGLF-ASTRA (see Fig.~\ref{fig:pressure_em_exb}). The TGLF simulations have been performed using the SAT1geo saturation rule \cite{Staebler_PoP_2016,Staebler_PPCF_2020}. The time-averaged fluxes are shown in Fig.~\ref{fig:flux_em_exb}.
Although the GENE standalone results are close to the volume integral of the injected sources for $\rho_{tor} \approx [0.3 - 0.5]$, an excellent agreement is obtained after running GENE-Tango for 29 iterations, each covering $t = 150c_s/a$. The overall GENE run time is $11.7$ms. The Tango relaxation coefficients acting on the plasma profiles and turbulent fluxes are fixed, respectively, to the $\alpha_p = 0.15$ and $\alpha_q = 0.4$.

\begin{figure*}
\begin{center}
\includegraphics[scale=0.40]{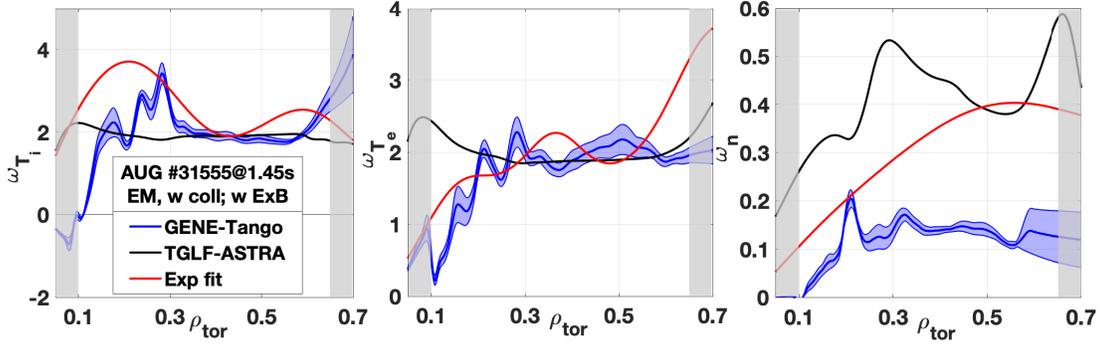}
\par\end{center}
\caption{Comparison of the a) ion, b) electron logarithmic temperatures and c) density gradients computed by TGLF-ASTRA (black), GENE-Tango averaged over the last five iterations (blue) and over the numerical fit of the experimental data (red). The shaded blue area represents the fluctuations of the different GENE-Tango profiles over the last five iterations. The gray areas indicate the locations of the buffer regions employed in the GENE simulations.}
\label{fig:gradients_em_exb}
\end{figure*}

The steady-state plasma profiles - averaged over the last five GENE-Tango iterations - are compared in Fig.~\ref{fig:pressure_em_exb} with those obtained with TGLF-ASTRA and the experimental measurements. While the electron temperature was reconstructed by combining two Thomson scattering systems (respectively one in the core while the other in the plasma edge) \cite{Kurzan_RSI_2011} and the electron cyclotron emission (ECE) system \cite{Rathgeber_PPCF_2012}, the plasma (electron) density with the lithium beam emission spectroscopy diagnostic \cite{Wolfrum_RSI_1993,Fischer_PPCF_2008} and the deuterium-cyanide-nitrogen laser interferometer \cite{Gehre_IJI_1984}. The charge exchange recombination spectroscopy systems was used for the ion temperature. To keep the separatrix electron temperature at $\sim 100$eV (typical value at ASDEX Upgrade) \cite{Neuhauser_PPCF_2002}, the electron temperature and density were shifted together. The dotted continuous red line in Fig.~\ref{fig:pressure_es} represents a fit over the experimental data. The fit was performed using a third order polynomial for the temperatures, and the two-line fit \cite{Schneider_PPCF_2012} for the density. Further details can be found in Ref.~\cite{Luda_NF_2021}.
To better quantify the differences among the plasma profiles and experiment, we add in Fig.~\ref{fig:gradients_em_exb} the logarithmic temperature and density profiles.

A first striking observation from Fig.~\ref{fig:pressure_em_exb} is the significant peaking of the ion temperature profiles - with respect to the TGLF-ASTRA profiles that remains rather flat - at $\rho_{tor} = 0.3$. This is seen more clearly when looking at the logarithmic ion temperature gradient, that exhibits a localized increase for $\rho_{tor} = [0.2 - 0.3]$ which is consistent with the experimental measurements. Nevertheless, the steady-state ion temperature profile computed with GENE-Tango is still under-predicted for $\rho_{tor}<0.2$. This is likely to be caused by the absence of supra-thermal particles in the GENE simulations. Energetic particles are known to stabilize - when properly optimized - ITG turbulence strongly, thus possibly leading to a further peaking of the ion temperature on-axis \cite{Citrin_PRL_2013,DiSiena_NF_2018,DiSiena_NF_2019}. The radial region $\rho_{tor} = [0 - 0.3]$ is the fraction of the plasma with the largest supra-thermal particle concentration in this ASDEX Upgrade discharge, due to the on-axis NBI heating scheme used in the experiments. The impact of supra-thermal particles on these results will be analyzes in the near future. This required some non trivial changes in the interface between GENE and Tango and will be performed in the near future.


Moreover, we observe from Fig.~\ref{fig:pressure_em_exb} an excellent agreement on the electron temperature between the different models and the experimental measurements. The only minor deviations are observed close to the GENE left buffer region, possibly related to the procedure used in Tango to account for the GENE inner buffer (namely by setting the amplitude of the sources to zero in the inner buffer region).

Despite the excellent agreement obtained on the ion and electron temperature profiles for $\rho_{tor} = [0.2 - 0.7]$, we notice that GENE-Tango systematically under-predicts the plasma density for this ASDEX Upgrade discharge with respect to the experimental measurements. This might be either related to missing physical effects in the GENE-Tango calculations or to an under-prediction of the plasma density at the GENE-Tango right boundary condition due to the large experimental uncertainties. 

\subsection{Improved agreement on the plasma density with boundary density value fixed by IDA} \label{IDA_sec}

\begin{figure*}
\begin{center}
\includegraphics[scale=0.40]{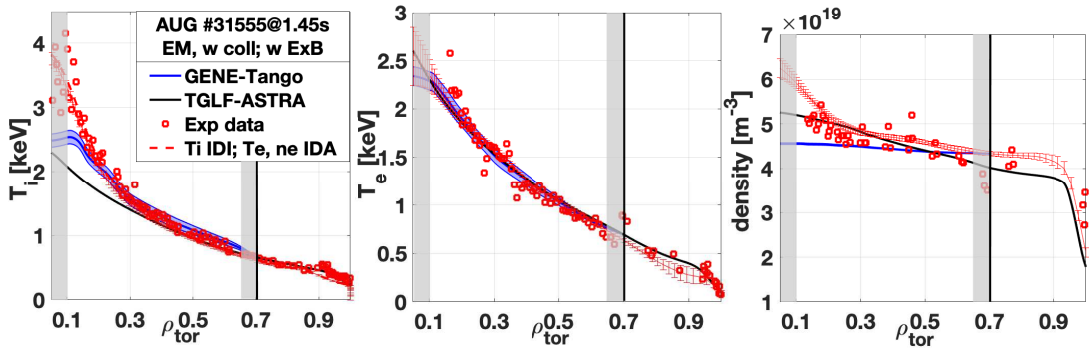}
\par\end{center}
\caption{Comparison of the a) ion, b) electron temperatures and c) density computed by TGLF-ASTRA (black), GENE-Tango averaged over the last five iterations (blue) and experimental measurements (red). The main ion temperature profile (and error bars) is obtained from IDI \cite{Fischer_FST_2020}, the electron temperature and density profiles (and error bars) from IDA \cite{Fischer_FSC_2010}. The shaded blue area represents the fluctuations of the different GENE-Tango profiles over the last five iterations. The vertical black line delimits the right boundary of the GENE-Tango radial  domain. The gray areas indicate the locations of the buffer regions employed in the GENE simulations.}
\label{fig:pressure_IDA}
\end{figure*}

In the previous section, we demonstrate that GENE-Tango can reproduce quantitatively the experimental temperature profiles of the ASDEX Upgrade discharge $\#31555$ at $t = 1.45$s, showing a pronounced peaking of the ion temperature not captured by TGLF-ASTRA. However, the density profile computed by GENE-Tango did not recover the experimental measurements, suggesting that the value of the plasma density at the GENE-Tango outer boundary was not consistent with the GENE turbulent levels. The choice of the plasma density (and temperatures) at the GENE-Tango outer boundary strongly influences the whole density profile at the inner core region. By looking at Fig.~\ref{fig:pressure_em_exb} we observe large uncertainties on the experimental measurements of the plasma density especially at $\rho_{tor} = 0.7$, thus making an accurate evaluation of the plasma density at the boundary particularly challenging. To assess whether an improved agreement with the experimental density profile could be achieved with a different choice of the boundary condition, we performed GENE-Tango simulations with a different value of the plasma density at $\rho_{tor} = 0.7$. The ion and electron temperature values at the GENE-Tango outer boundary are unchanged. The value of the density at $\rho_{tor} = 0.7$ has been fixed to that of the IDA \cite{Fischer_FSC_2010} density profile (red line in Fig.~\ref{fig:pressure_IDA}). It differs with respect to those employed by TGLF-ASTRA by $\approx 8\%$ (compatible with the experimental error bars), which was obtained with the integrated model based on engineering parameters (IMEP) introduced in Ref.~\cite{Luda_NF_2020}.

To reduce the computational cost of the simulation, we initialized the GENE-Tango iterations by running a GENE stand-alone simulation with the steady-state temperature profiles obtained with the TGLF-ASTRA boundaries (Fig.~\ref{fig:pressure_em_exb}) and by re-scaling the converged plasma density with the new boundary value. With this setup we reached the new solution for the plasma profiles with $10$ GENE-Tango iterations, each of $t = 150 c_s /a$. The steady-state profiles are shown in Fig.~\ref{fig:pressure_IDA}.

By looking at Fig.~\ref{fig:pressure_IDA} we observe not only an excellent agreement on the ion and electron temperature profiles, but also an improved agreement for the plasma density, which well sits inside the experimental measurements in $\rho_{tor} = [0.3 - 0.7]$. However, we still notice an under-prediction of the density profile in $\rho_{tor} = [0 - 0.2]$. As mentioned above, this result might be due to the absence of supra-thermal particles in the GENE simulations, which are known to strongly suppress heat and particle transport (see e.g., \cite{DiSiena_PoP_2018}), thus leading to increased core gradients. The impact of fast particles on the evolution of the plasma profiles computed by GENE-Tango will be addressed in the near future.

It is worth mentioning here that the GENE-Tango simulations presented in this and the previous Section lasted, respectively, 24h and 72h on 16 IBM POWER9 AC922 nodes, each with 4 Nvidia Volta V100 GPUs on Marconi100. Although the computational cost of the GENE-Tango simulations is greater than reduced models like TGLF-ASTRA (a few hours on a single CPU \cite{Luda_NF_2020}), it still allows the routine use of GENE-Tango for discharge analyses. A detailed discussion on the GENE-Tango speedup and extrapolations to ITER can be found in Section \ref{sec11}.

\section{Impact of different physical effects on plasma pressure profiles} \label{impact}

In the previous Section, an excellent agreement between the GENE-Tango profiles and the experimental measurements was found even in regimes where TGLF-ASTRA under-predicts the peaking of the on-axis ion temperature profile. Given the moderate computational resources required to run GENE-Tango up to the transport time-scale, in this Section we perform several GENE-Tango simulations that retain different physical effects in the turbulence calculations. These analyses are essential to identify the mechanism responsible for the increased on-axis ion temperature and, possibly, provide insights into improving reduced turbulence models.

Therefore, the impact of collisions, toroidal plasma rotation and electromagnetic effects on the evolution of the plasma profiles is studied with GENE-Tango simulations of the ASDEX Upgrade discharge $\# 31555$ at $t = 1.45$s. To compare more easily the GENE-Tango profiles with TGLF-ASTRA, we employ the same boundary values on temperatures and density used in TGLF-ASTRA. However, we note - as shown in Section \ref{IDA_sec} - that GENE-Tango could achieve an improved agreement with the experiment when fixing the plasma density at $\rho_{tor} = 0.7$ to the value obtained with IDA. The extrapolation regions used in Tango on the particle and heat fluxes computed by GENE are kept the same for each of the different cases studied here in the final GENE-Tango iterations. These are, respectively, $x_{Ti} = [0.52 - 0.56]$, $x_{Te} = [0.52 - 0.56]$ and $x_{n} = [0.45 - 0.49]$ for ion, electron temperatures and plasma density.

\subsection{Electrostatic GENE-Tango simulation with collisions and without external $E \times B$ shear}  \label{sec6}

\begin{figure*}
\begin{center}
\includegraphics[scale=0.40]{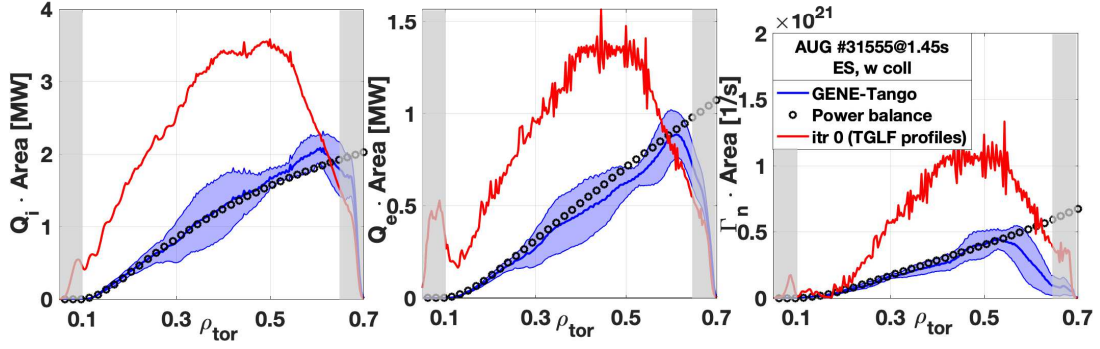}
\par\end{center}
\caption{Time-averaged radial profile of the a) ion, b) electron heat fluxes in MW and c) particle flux in $1/s$ corresponding to the stand-alone GENE simulation with the initial TGLF-ASTRA profiles (red) and the last 5 GENE-Tango iterations (blue). The shaded blue area represents the fluctuations of the turbulent fluxes over the last five iterations. The gray areas denote the buffer regions and the black circles the volume integral of the injected particle and heat sources.}
\label{fig:flux_es}
\end{figure*}

\begin{figure*}
\begin{center}
\includegraphics[scale=0.40]{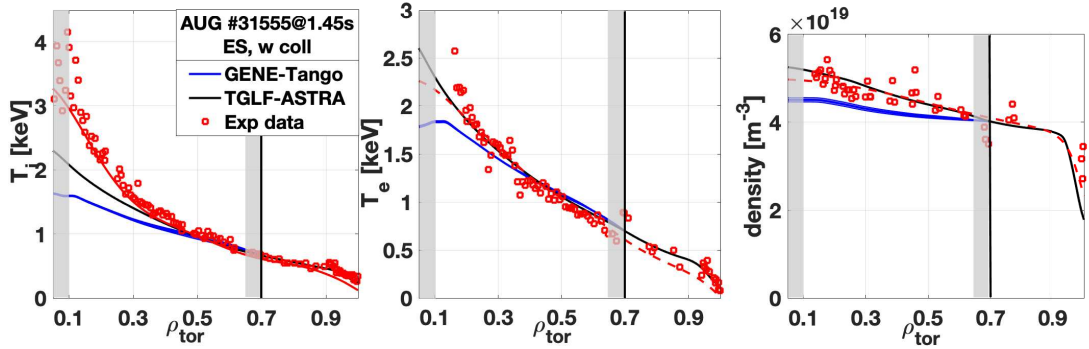}
\par\end{center}
\caption{Comparison of the a) ion, b) electron temperatures and c) density computed by TGLF-ASTRA (black), GENE-Tango averaged over the last five iterations (blue) and experimental measurements (red). The shaded blue area represents the fluctuations of the different GENE-Tango profiles over the last five iterations. The vertical black line delimits the right boundary of the GENE-Tango radial domain. The gray areas indicate the locations of the buffer regions employed in the GENE simulations.}
\label{fig:pressure_es}
\end{figure*}

We begin these analyses in the electrostatic limit ($\beta_e(x_0) = 8 \pi n(x_0)T_e(x_0) / B_0^2 = 1e-4$ with $x_0 = 0.375$) with collisions modeled with a linearized Landau-Boltzmann operator. The numerical setup and physical parameters are the same as the ones summarized in Section \ref{sec1}. We initialize the iterations with Tango by running a GENE standalone simulation using the TGLF-ASTRA profiles. The results are illustrated in Fig.~\ref{fig:flux_es}. Large turbulence fluxes are observed in all the different channels, which are not compatible with the volume average of the injected sources, thus suggesting that the TGLF-ASTRA profiles cannot be sustained in the GENE-Tango electrostatic simulations with collisions. Starting from these reference results, an excellent agreement between the GENE fluxes and the experimental power balance is achieved after $20$ GENE-Tango iterations, corresponding to an overall GENE run-time of $8.1$ms. This is shown in Fig.~\ref{fig:flux_es} where the heat and particle fluxes obtained over the last $5$ iterations are compared with the volume integral of the injected sources. For these analyses, the Tango relaxation coefficients acting on the plasma profiles and turbulent fluxes are kept constant, respectively, to the values $\alpha_p = 0.2$ and $\alpha_q = 0.4$. The GENE run time (for each iteration) was set to $150 c_s / a$.
The temperatures and density computed by Tango over the last $5$ GENE-Tango iterations are compared in Fig.~\ref{fig:pressure_es} with the TGLF-ASTRA profiles and the experimental measurements. 

These results show a good agreement between the electron temperature profiles. However, GENE-Tango predicts a more pronounced flattening of the ion temperature for $\rho_{tor}<0.4$ compared to TGLF-ASTRA, thus making $T_i$ further deviating with respect to the experimental measurements. While the electron temperatures follow the experimental measurements closely, we notice that both the electrostatic GENE-Tango ion temperature and the TGLF-ASTRA one do not exhibit the on-axis peaking observed in the experiment. 
Moreover, Fig.~\ref{fig:pressure_es} shows that the density profile computed with GENE-Tango deviates respect to the TGLF-ASTRA profile and the experimental measurements. In particular, GENE-Tango predicts a rather flat profile not in agreement with the experiment and TGLF-ASTRA. This result is consistent with Fig.~\ref{fig:pressure_em_exb} and provide further evidence indicating that the value of the plasma density on the right boundary of the GENE-Tango simulation (fixed to the TGLF-ASTRA one) is not fully consistent with the transport levels computed by GENE. The impact of the plasma density at the GENE-Tango outer boundary was discussed in detail in Section \ref{IDA_sec}.

\begin{figure*}
\begin{center}
\includegraphics[scale=0.40]{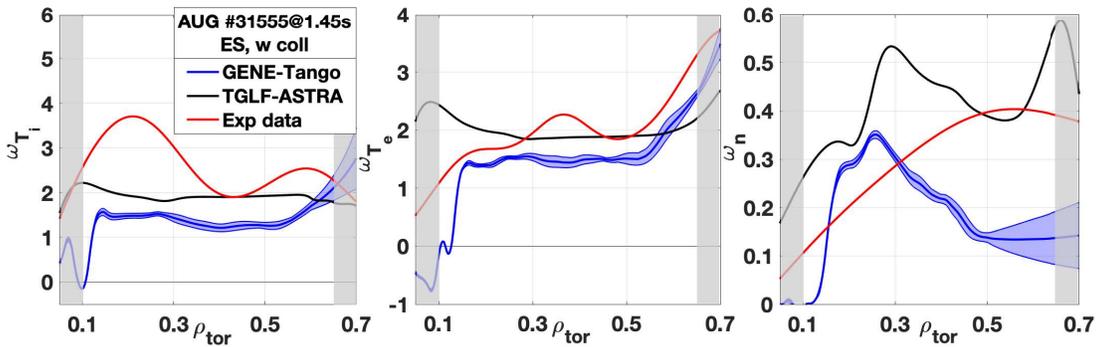}
\par\end{center}
\caption{Comparison of the a) ion, b) electron logarithmic temperatures and c) density gradients computed by TGLF-ASTRA (black), GENE-Tango averaged over the last five iterations (blue) and over the numerical fit of the experimental measurements (red). The shaded blue area represents the fluctuations of the different GENE-Tango profiles over the last five iterations. The gray areas indicate the locations of the buffer regions employed in the GENE simulations.}
\label{fig:gradients_es}
\end{figure*}

To better quantify variations among the different profiles, we show in Fig.~\ref{fig:gradients_es} the logarithmic temperature and density gradients. The logarithmic gradient of the experimental data has been computed on the fitted line. These analyses confirm the good agreement between GENE-Tango and TGLF-ASTRA on the electron temperature. 
More pronounced variations 
are observed for $\omega_{T_i}$ and $\omega_{n}$ with reduced gradients predicted by Tango in most of the radial domain considered. 

\subsection{Electrostatic GENE-Tango simulation with collisions and external $E \times B$ shear}  \label{sec7}

The under-prediction of the on-axis temperature with TGLF-ASTRA is often attributed to electromagnetic and supra-thermal particle effects on plasma turbulence that are not fully captured yet by reduced turbulence models, such as TGLF \cite{Doerk_NF_2017,Mantica_PPCF_2019,Reisner_NF_2020,Luda_NF_2021}. However, these codes have been shown recently to qualitatively recover the experimental peaking on $T_i$ by artificially increasing the amplitude of the toroidal rotation (called $v_{tor}$ in the remaining of this Section) above the experimental uncertainties \cite{Reisner_NF_2020}. Nevertheless, high-fidelity gyrokinetic codes exhibit only a weak dependence in the plasma core on the turbulent fluxes with the toroidal rotation (for experimentally relevant amplitudes of the flow shear), despite the strong effect observed on reduced turbulence codes \cite{Citrin_PRL_2013,Garcia_NF_2015,Doerk_PPCF_2016}. The role of the toroidal rotation on the evolution of the plasma profiles computed with GENE-Tango will be addressed in detail in the present Section and in Section \ref{sec9}.

Toroidal rotation in magnetic confinement devices leads to a radial electric field and hence to a finite $E \times B$ plasma flow, typically reducing the turbulence correlation lengths and outward particle and energy fluxes \cite{Burrell_PoP_1997,Connor_PPCF_2004}. An efficient way of inducing these large-scale plasma flows in experiments is via neutral beam injection. Given the non-negligible NBI heating power used in the ASDEX Upgrade discharge $\#31555$ at $t=1.45$s, we investigate here the role of the $E\times B$ shear on the $T_i$ peaking observed in the experiment with gyrokinetic simulations up to the transport time scale with GENE-Tango. The profile of the toroidal rotation (not evolved in Tango) used in this Section is the same as the one shown in Fig.~\ref{fig:vtor}.
\begin{figure*}
\begin{center}
\includegraphics[scale=0.40]{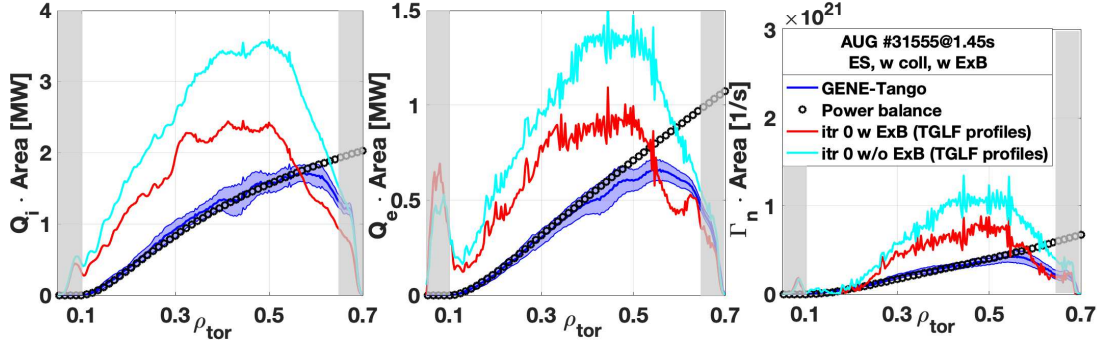}
\par\end{center}
\caption{Time-averaged radial profile of the a) ion, b) electron heat fluxes in MW and c) particle flux in $1/s$ corresponding to the stand-alone GENE simulation with the initial TGLF-ASTRA profiles, respectively, with (red) and without (cyan) toroidal rotation and the last 5 GENE-Tango iterations (blue). The shaded blue area represents the fluctuations of the turbulent fluxes over the last five iterations. The gray areas denote the buffer regions and the black circles the volume integral of the injected particle and heat sources.}
\label{fig:flux_es_exb}
\end{figure*}

To begin the GENE-Tango iterations, we run at first a GENE standalone simulation to convergence using the TGLF-ASTRA temperature and density profiles. The time-averaged turbulent fluxes obtained with GENE are illustrated in Fig.~\ref{fig:flux_es_exb} (red line) and compared with the initial GENE simulation without external toroidal rotation (cyan line) (Fig.~\ref{fig:flux_es}). Fig.~\ref{fig:flux_es_exb} shows (with respect to the electrostatic GENE-Tango results without external toroidal rotation) that the inclusion of the external toroidal rotation leads to a turbulence stabilization in all the different channels. While only a mild reduction in heat and particle fluxes is observed for $\rho_{tor}<0.3$ (roughly $\sim 15\%$), we observe a progressive enhancement in the outer regions (roughly $42\%$ at $\rho_{tor} = 0.6$). This turbulence stabilization brings the initial GENE turbulent fluxes closer to the volume integral of the injected sources (black dots of Fig.~\ref{fig:flux_es_exb}). However, the dependence of the ion heat flux with $v_{tor}$ is not consistent with the experimental measurements of $T_i$. In particular, the on-axis temperature profile peaking observed in the experiment would require a localized stabilization in the deep core regions. This is in contrast with the effect of $v_{tor}$ on the GENE fluxes observed in Fig.~\ref{fig:flux_es_exb} (localized more in the plasma edge).
\begin{figure*}
\begin{center}
\includegraphics[scale=0.40]{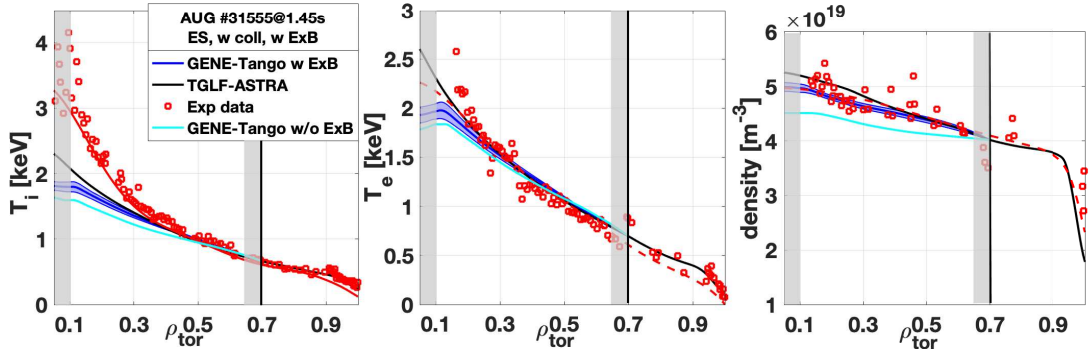}
\par\end{center}
\caption{Comparison of the a) ion, b) electron temperatures and c) density computed by TGLF-ASTRA (black), GENE-Tango with toroidal rotation averaged over the last five iterations (blue), GENE-Tango without toroidal rotation (cyan) and experimental measurements (red). The shaded blue area represents the fluctuations of the different GENE-Tango profiles over the last five iterations. The vertical black line delimits the right boundary of the GENE-Tango radial domain. The gray areas indicate the locations of the buffer regions employed in the GENE simulations.}
\label{fig:pressure_es_exb}
\end{figure*}

\begin{figure*}
\begin{center}
\includegraphics[scale=0.40]{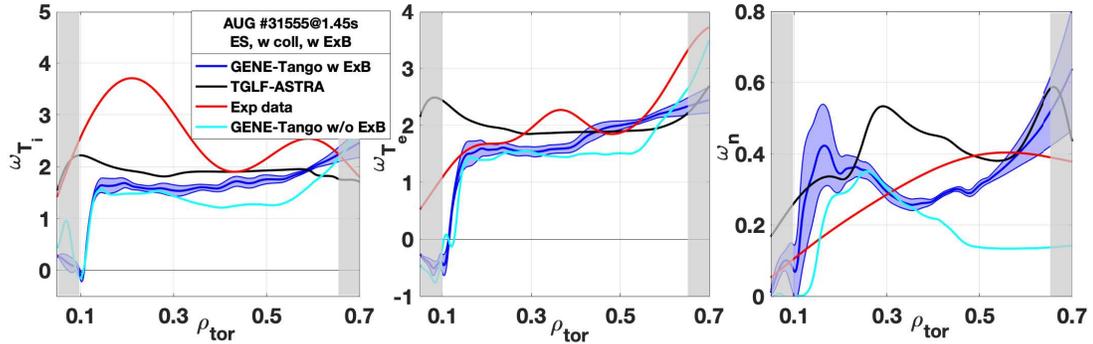}
\par\end{center}
\caption{Comparison of the a) ion, b) electron logarithmic temperatures and c) density gradients computed by TGLF-ASTRA (black), GENE-Tango with toroidal rotation averaged over the last five iterations (blue), GENE-Tango without toroidal rotation (cyan) and over the numerical fit of the experimental measurements (red). The shaded blue area represents the fluctuations of the different GENE-Tango profiles over the last five iterations. The gray areas indicate the locations of the buffer regions employed in the GENE simulations.}
\label{fig:gradients_es_exb}
\end{figure*}

Starting from the initial fluxes of Fig.~\ref{fig:flux_es_exb}, we perform 15 GENE-Tango iterations before reaching the steady-state solution where the GENE fluxes match the power balance in all turbulent channels. For each iteration, GENE has been run up to $t = 150c_s /a$, thus leading to an overall GENE run-time of $6.1$ms. The Tango relaxation coefficients acting on the plasma profiles and turbulent fluxes are kept constant, respectively, to the value $\alpha_p = 0.2$ and $\alpha_q = 0.4$. The converged temperature and density profiles computed by Tango - averaged over the last five GENE-Tango iterations - are illustrated in Fig.~\ref{fig:pressure_es_exb} and compared with the TGLF-ASTRA ones, the electrostatic GENE-Tango profiles without external toroidal rotation and the experimental measurements. We also add the logarithmic temperature and density profiles in Fig.~\ref{fig:gradients_es_exb} to better quantify the differences among the models and the experiment.

Fig.~\ref{fig:pressure_es_exb} shows an excellent agreement between the GENE-Tango and TGLF-ASTRA profiles with only minor variations close to the GENE left buffer region. As discussed already in Section \ref{sec5}, the flattening of the GENE-Tango profiles at the left boundary might be related to the procedure used in Tango to account for the GENE inner buffer (namely by setting the amplitude of the sources to zero in the inner buffer region).

The most considerable effect of the toroidal rotation is observed on the plasma density. While it was under-predicted in the GENE-Tango simulations of Fig.~\ref{fig:pressure_es}, it is fully recovered when the toroidal plasma rotation is retained in the GENE simulations. In particular, we notice a significant peaking of the density profile primarily located in the outer regions, thus improving the agreement of the GENE-Tango density profile with the experimental measurements and TGLF-ASTRA. These findings are in agreement with the flux-tube results of Ref.~\cite{Garcia_PPCF_2019} suggesting that the $E\times B$ shear might enhance inward particle pinch.

Consistently with the results of Fig.~\ref{fig:flux_es_exb}, we observe that the effect of toroidal rotation on the temperature profiles is localized in the outer core regions, i.e.,~$\rho_{tor} = [0.4 - 0.6]$. This is illustrated clearly in Fig.~\ref{fig:gradients_es_exb} showing the comparison between the logarithmic temperature gradients obtained with and without toroidal rotation.

However, the ion temperature profile still does not exhibit the pronounced on-axis peaking characteristic of the ASDEX Upgrade discharge $\#31555$ at $t = 1.45$s. In agreement with previous flux-tube results, these findings suggest that the toroidal rotation can hardly explain the $T_i$ peaking in gyrokinetic codes in the plasma core \cite{Citrin_PRL_2013,Garcia_NF_2015}.

As it will be shown and discussed in Section \ref{sec9} the increase of the ion temperature on-axis experienced in experiments with large external heating is most likely due to electromagnetic and fast particle effects \cite{Citrin_PRL_2013,Whelan_PRL2018,DiSiena_NF_2019}, which are known to be strongly underestimated by reduced turbulence models. This interpretation is consistent with the results of Fig.~\ref{fig:pressure_es_exb}, showing that electrostatic GENE-Tango simulations retaining toroidal rotation are sufficient to reproduce the temperature and density profiles computed by TGLF-ASTRA, suggesting that the model used in TGLF is essentially electrostatic.

\subsection{Electrostatic GENE-Tango simulation without collisions and without external $E \times B$ shear}  \label{sec8}

\begin{figure*}
\begin{center}
\includegraphics[scale=0.40]{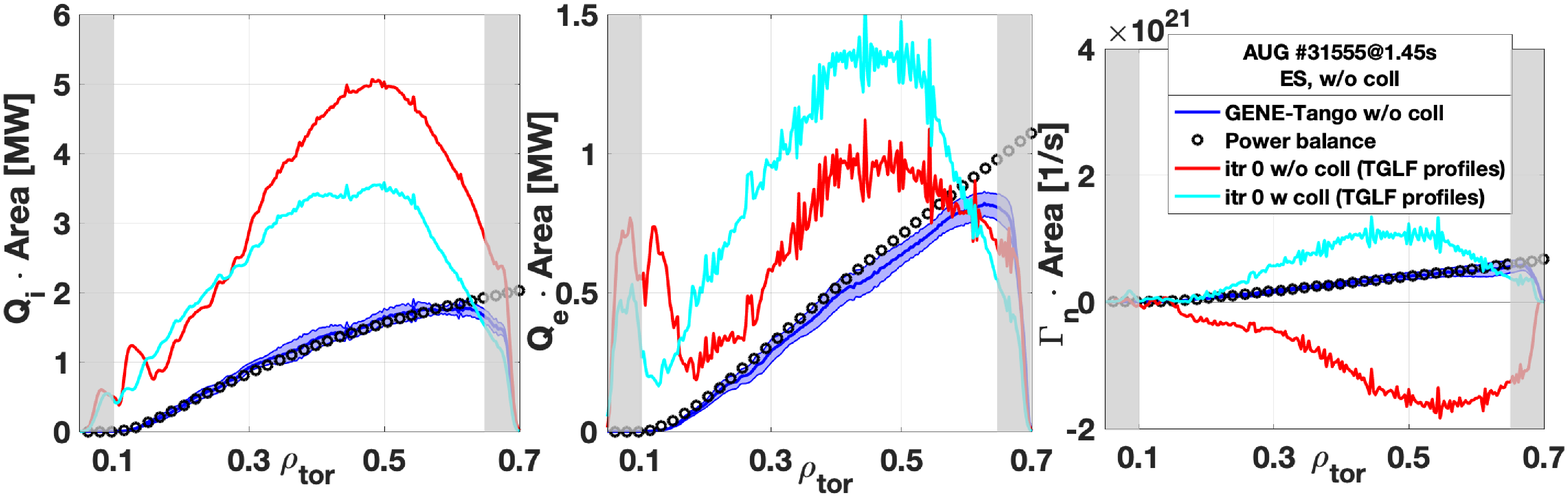}
\par\end{center}
\caption{Time-averaged radial profile of the a) ion, b) electron heat fluxes in MW and c) particle flux in $1/s$ corresponding to the stand-alone GENE simulation with the initial TGLF-ASTRA profiles, respectively, without (red) and with collisions (cyan) and the last 5 GENE-Tango iterations (blue). The shaded blue area represents the fluctuations of the turbulent fluxes over the last five iterations. The gray areas denote the buffer regions and the black circles the volume integral of the injected particle and heat sources.}
\label{fig:flux_nocoll}
\end{figure*}

The effect of collisions on the GENE electrostatic fluxes and plasma profiles is investigated within this Section. In particular, we perform electrostatic GENE-Tango simulations neglecting collisions in the turbulence calculations. To initialize the GENE-Tango iterations, we carry out a reference standalone GENE simulation with the profiles computed by TGLF-ASTRA. These initial turbulent fluxes are illustrated in Fig.~\ref{fig:flux_nocoll} by the continuous red line.
When comparing these reference results with those obtained in the electrostatic GENE-Tango results without toroidal rotation (Fig.~\ref{fig:flux_es}), we note an overall increase of the ion heat flux in the absence of collisions by roughly $40\%$ at $\rho_{tor}= 0.5$. On the other hand, the electron heat flux is reduced by $25\%$ at $\rho_{tor} = 0.5$. The particle flux undergoes the most violent changes, exhibiting large inward fluxes in the whole radial region of interest. This result is well consistent with the theoretical model derived in Ref.~\cite{Fable_PPCF_2009}. In particular, collisions alter the pure convective term \cite{Weiland_NF_1989,Garbet_PRL_2003} in the quasi-linear decomposition of the particle flux by adding a positive-defined contribution in ITG turbulence \cite{Angioni_PoP_2009,Angioni_NF_2012}. This term typically balances - together with the diagonal diffusion (proportional to the logarithmic density gradient) - the inward thermo-diffusion term in ITG regimes, leading to a net outward flux (as shown in Fig.~\ref{fig:flux_es}). However, the outward collisional contribution vanishes in collisionless plasmas, resulting in an inward particle flux if the diffusion term is not large enough to balance this negative contribution. 

Although the initial profiles are far from matching the volume integral of the injected particle and energy sources, an excellent agreement is obtained in Fig.~\ref{fig:flux_nocoll} after running GENE-Tango for $30$ iterations, each covering $150 c_s / a$. This corresponds to an overall GENE run-time of $12.7$ms. The Tango relaxation coefficients acting on the plasma profiles and turbulent fluxes are kept constant, respectively, to the values $\alpha_p = 0.1$ and $\alpha_q = 0.4$. 
\begin{figure*}
\begin{center}
\includegraphics[scale=0.40]{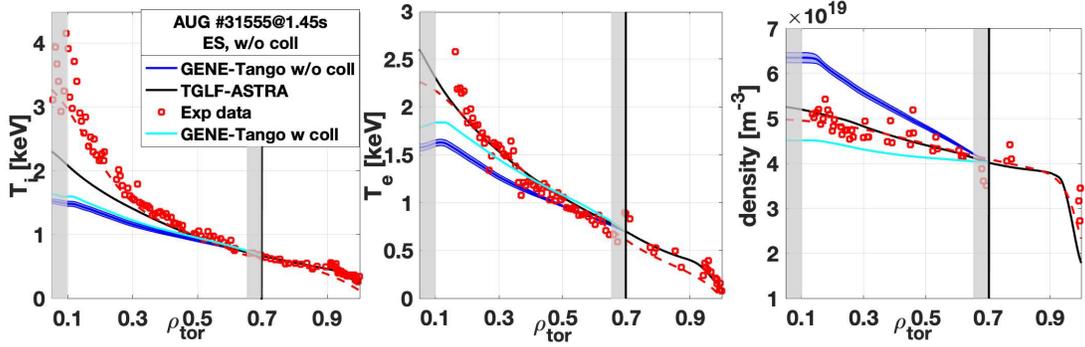}
\par\end{center}
\caption{Comparison of the a) ion, b) electron temperatures and c) density computed by TGLF-ASTRA (black), GENE-Tango without collisions averaged over the last five iterations (blue), GENE-Tango with collisions (cyan) and experimental measurements (red). The shaded blue area represents the fluctuations of the different GENE-Tango profiles over the last five iterations. The vertical black line delimits the right boundary of the GENE-Tango radial domain. The gray areas indicate the locations of the buffer regions employed in the GENE simulations.}
\label{fig:pressure_nocoll}
\end{figure*}

The converged temperatures and density computed by Tango over the last $5$ GENE-Tango iterations are compared in Fig.~\ref{fig:pressure_nocoll} with the TGLF-ASTRA profiles, the experimental measurements, and the GENE-Tango profiles obtained retaining the effect of collisions. Consistently with the reference results (Fig.~\ref{fig:flux_nocoll}) - showing an increase in the ion heat flux in the absence of collisions -, we note that the ion temperature is further flattened respect to profile including collisions to match the power balance. Moreover, we observe a significant peaking in the plasma density. This is a consequence of the large inward particle flux contribution obtained without collisions. In particular, a positive particle flux - fixed by the external sources - can be matched only by enhancing the positive-defined diagonal diffusion term, hence by increasing the logarithmic density gradient. This dependence of the plasma density with collisionality is consistent with the theoretical results of Ref.~\cite{Angioni_PRL_2003,Angioni_PoP_2009} and the experimental evidence observed in different magnetic confinement devices \cite{Angioni_PRL_2003,Angioni_PoP_2003,Weisen_NF_2005,Greenwald_NF_2007,Angioni_NF_2007,Takenaga_NF_2008}.

\begin{figure*}
\begin{center}
\includegraphics[scale=0.40]{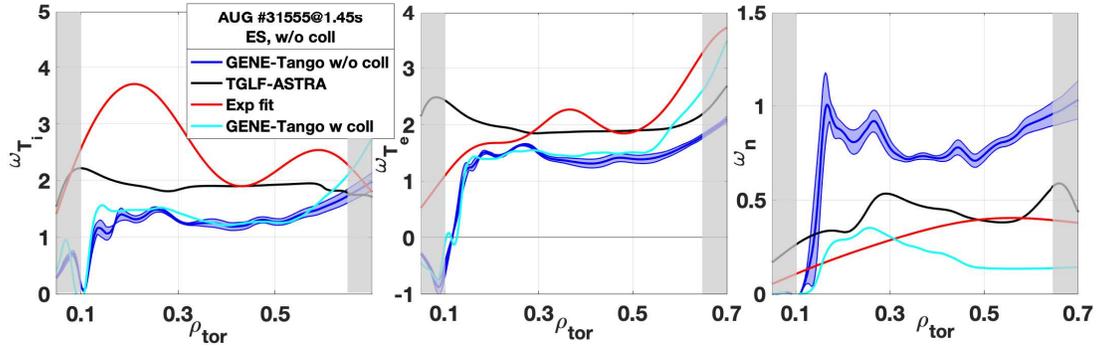}
\par\end{center}
\caption{Comparison of the a) ion, b) electron logarithmic temperatures and c) density gradients computed by TGLF-ASTRA (black), GENE-Tango with collisions averaged over the last five iterations (blue), GENE-Tango with collisions (cyan) and over the numerical fit of the experimental measurements (red). The shaded blue area represents the fluctuations of the different GENE-Tango profiles over the last five iterations. The gray areas indicate the locations of the buffer regions employed in the GENE simulations.}
\label{fig:gradients_nocoll}
\end{figure*}

Another observation from Fig.~\ref{fig:pressure_nocoll}, is the flattening of the electron temperature profile respect to the results of Fig.~\ref{fig:pressure_es}. This is due to the enhancement of trapped-electron-mode (TEM) turbulence with the increased density gradient. For completeness, we show in Fig.~\ref{fig:gradients_nocoll} a comparison of the logarithmic temperature and density gradient for the different cases studied within this section. 
Given the central role played by collisions in setting the plasma density profile, more accurate operators (such as the Sugama model \cite{Sugama_PoP_2009} and the exact Fokker-Planck operator \cite{Qingjiang_PoP_2020}) will be employed in the near future.

\subsection{Electromagnetic GENE-Tango simulation with collisions and without external $E \times B$ shear}  \label{sec9}

In the previous Sections, we have investigated the effect of toroidal rotation and collisions on the electrostatic turbulent fluxes and bulk profiles. The inclusion of external $E \times B$ shear in the GENE simulations leads to an excellent agreement between the temperature and density profiles computed by GENE-Tango and TGLF-ASTRA. Collisions are found to be essential to model the particle flux correctly and hence to set the plasma density. A minor impact has been identified, on the other hand, on the bulk species temperatures. Interestingly, both the TGLF-ASTRA and GENE-Tango (electrostatic) profiles with/without toroidal rotation do not capture the on-axis peaking of the ion temperature observed in the experiment. The computed profiles start to deviate from the measurements for $\rho_{tor}<0.4$. This is the plasma region where electromagnetic effects are expected to become more relevant. In such electromagnetic regimes, an increasing number of studies are showing that presently available reduced turbulence models fail to fully capture electromagnetic effects on plasma micro-turbulence, often leading in gyrokinetic codes to a significant turbulence stabilization.

\begin{figure*}
\begin{center}
\includegraphics[scale=0.40]{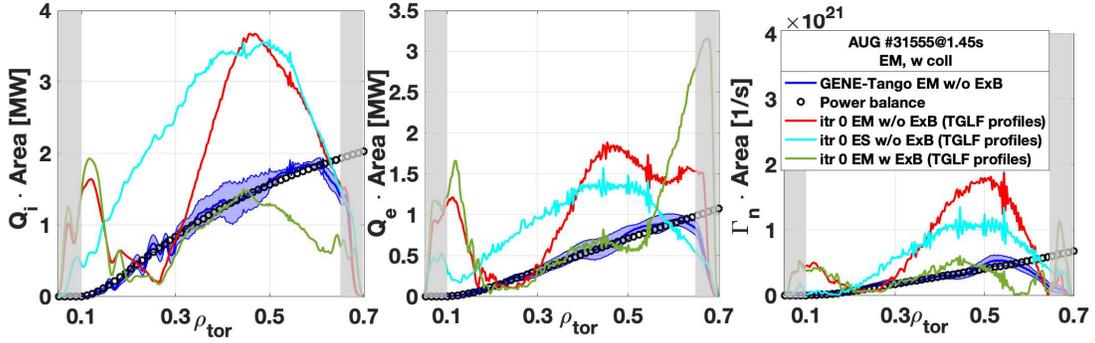}
\par\end{center}
\caption{Time-averaged radial profile of the a) ion, b) electron heat fluxes in MW and c) particle flux in $1/s$ corresponding to the stand-alone GENE simulation with the initial TGLF-ASTRA profiles, respectively, with electromagnetic effects and no toroidal rotation (red), in the electrostatic limit and no toroidal rotation (cyan) and with electromagnetic effects and toroidal rotation (green) and the last 5 GENE-Tango iterations (blue). The shaded blue area represents the fluctuations of the turbulent fluxes over the last five iterations. The gray areas denote the buffer regions and the black circles the volume integral of the injected particle and heat sources.}
\label{fig:flux_em}
\end{figure*}

In this Section, we assess the role of finite beta on the bulk profiles by performing GENE-Tango electromagnetic simulations with collisions. We begin by running a GENE standalone simulation to convergence, starting from the reference TGLF-ASTRA profiles. The time-averaged turbulent fluxes are shown in Fig.~\ref{fig:flux_em} by the continuous red line. Compared to the reference GENE electrostatic simulation (cyan line), we note that the inclusion of electromagnetic effects strongly affects the turbulent fluxes in all different channels. To begin with, we observe a significant increase of heat and particle fluxes at $\rho_{tor} = 0.1$, related to the destabilization of an electromagnetic mode with the TGLF-ASTRA profiles (not shown here). Moreover, while no quantitative differences are found on the ion heat flux for $\rho_{tor} > 0.4$, the electron heat and particle fluxes increase by roughly $\sim 40 \%$ at $\rho_{tor} = 0.5$. This is consistent with previous flux-tube findings showing that the inclusion of electromagnetic effects in gyrokinetic simulations without supra-thermal particles might destabilize the electron heat flux \cite{DiSiena_JPP_2021}. However, and more importantly, we observe a significant turbulence suppression in the radial region $\rho_{tor} = [0.15 - 0.35]$, which coincides with the radial locations where the ion temperature profile peaks in the experiments but not in the TGLF-ASTRA and electrostatic GENE-Tango profiles.

To further analyze the impact of the toroidal plasma rotation on the GENE-Tango results (which might affect the turbulent fluxes differently in electromagnetic regimes), the turbulent fluxes obtained in the electromagnetic GENE simulation retaining $v_{tor}$ (see Fig.~\ref{fig:flux_em_exb}) are added to Fig.~\ref{fig:flux_em} (green line). While minor differences are observed on the heat and particle fluxes in the deep core regions (i.e., $\rho_{tor} <0.35$), a significant turbulence suppression is found in all channels for $\rho_{tor}>0.4$ for a finite toroidal rotation. The electron heat flux decreases with respect to the case with toroidal rotation at $\rho_{tor} \approx 0.65$. These results are consistent with the ones shown in Fig.~\ref{fig:flux_es_exb}, showing an increased effect of a finite $E \times B$ in the plasma edge.

\begin{figure*}
\begin{center}
\includegraphics[scale=0.40]{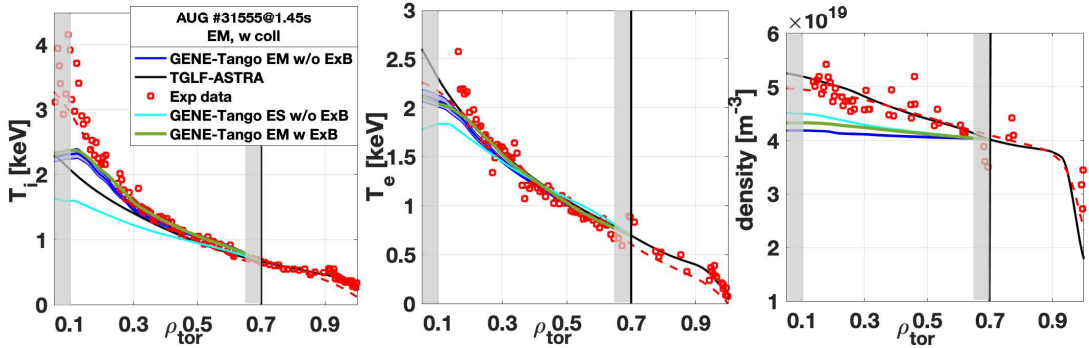}
\par\end{center}
\caption{Comparison of the a) ion, b) electron temperatures and c) density computed by TGLF-ASTRA (black), GENE-Tango electromagnetic without toroidal rotation averaged over the last five iterations (blue), GENE-Tango electrostatic with no toroidal rotation (cyan), GENE-Tango electromagnetic with toroidal rotation (green) and experimental measurements (red). The vertical black line delimits the right boundary of the GENE-Tango radial domain. The shaded blue area represents the fluctuations of the different GENE-Tango profiles over the last five iterations. The vertical black line delimits the right boundary of the GENE-Tango radial domain. The gray areas indicate the locations of the buffer regions employed in the GENE simulations.}
\label{fig:pressure_em}
\end{figure*}

\begin{figure*}
\begin{center}
\includegraphics[scale=0.40]{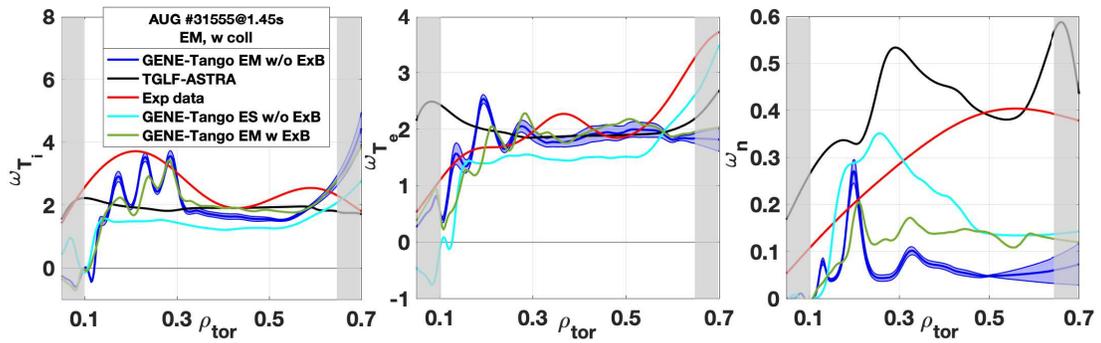}
\par\end{center}
\caption{Comparison of the a) ion, b) electron logarithmic temperatures and c) density gradients computed by TGLF-ASTRA (black), electromagnetic GENE-Tango with no toroidal rotation averaged over the last five iterations (blue), electrostatic GENE-Tango with no toroidal rotation (cyan), electromagnetic GENE-Tango with toroidal rotation (green) and over the numerical fit of the experimental measurements (red). The shaded blue area represents the fluctuations of the different GENE-Tango profiles over the last five iterations. The gray areas indicate the locations of the buffer regions employed in the GENE simulations.}
\label{fig:gradients_em}
\end{figure*}

Starting from these initial profiles, we performed 28 GENE-Tango iterations, each covering a time domain of $t = 150 c_s / a$ to reach steady-state. The overall GENE run time measures $11.7$ms. For these analyses the relaxation coefficients used in Tango for the plasma profiles and fluxes are respectively $\alpha_p = 0.15$ and $\alpha_q = 0.4$. The beta profile is computed self-consistently from the Tango profiles.
The results are illustrated in Fig.~\ref{fig:flux_em}, showing an excellent agreement between the GENE fluxes and the volume integral of the injected sources. The steady-state temperatures and density profiles computed by Tango over the last five iterations are depicted in Fig.~\ref{fig:pressure_em} and compared with the electrostatic ones (Fig.~\ref{fig:pressure_es}), the ones with electromagnetic and toroidal rotation (Fig.~\ref{fig:pressure_em_exb}), the TGLF-ASTRA profiles and the experimental measurements. To better visualize the differences among these profiles, we show in Fig.~\ref{fig:gradients_em} the logarithmic gradients as well.

We start by comparing the electrostatic (Fig.~\ref{fig:pressure_es}) with the electromagnetic GENE-Tango profiles. Interestingly, we observe a significant peaking of the ion temperature - with respect to the electrostatic GENE-Tango profile - starting from the radial location $\rho_{tor} = 0.3$. Its on-axis value increases from $T_i =1.6$keV to $T_i = 2.3$keV when electromagnetic effects are retained in the GENE simulations. Due to this pronounced peaking the electromagnetic GENE-Tango temperature profile shows an excellent agreement with the experimental measurements for $\rho_{tor}>0.2$. This can be observed also by looking at the logarithmic ion temperature gradient, which exhibits the characteristic increase at $\rho_{tor} = [0.2 - 0.3]$ of the experimental measurements. These findings provide numerical evidence showing that the peaking of the ion temperature observed in several discharges and still not fully captured by reduced models is likely to be caused by electromagnetic effects on plasma turbulence. As discussed in Section \ref{sec5}, the under-prediction of $T_i$ for $\rho_{tor}<0.2$ in GENE-Tango could be caused by the absence of supra-thermal particles in the GENE simulations. 

The electron temperature profile undergoes a mild flattening for $\rho_{tor} > 0.4$ and a minor peaking for $\rho_{tor} = 0.3$ (more evident in Fig.~\ref{fig:gradients_em}). This is consistent with the reference results of Fig.~\ref{fig:flux_em}. We note the excellent agreement of the electron temperature compared with the experimental measurements.

The density profile, on the other hand, further deviates from the experimental profile for the GENE-Tango case. In particular, its on-axis value reduces in the electromagnetic simulations. This result is in agreement with the theoretical predictions of Ref.~\cite{Hein_PoP_2010}, showing that electromagnetic fluctuations largely enhance the outward flux contribution of passing electrons in ITG dominated turbulence regimes, thus leading to an increase in the particle flux. To match the power balance on the particle flux - fixed only by the NBI scheme and particle refueling coming from the SOL neutrals -, Tango further flattens the density profile. The impact of finite beta on the density profile is particularly visible in Fig.~\ref{fig:gradients_em}, where the logarithmic density gradient is depicted.

When we compare the electromagnetic GENE-Tango profiles in Fig.~\ref{fig:pressure_em} obtained retaining and neglecting toroidal rotation, we observe no qualitative differences on the ion and electron temperatures. Both these profiles are found to quantitatively reproduce the experimental measurements up to $\rho_{tor} \approx 0.2$. To measure more clearly the (minor) impact of the toroidal plasma rotation of the temperature profiles we refer the reader to Fig.~\ref{fig:gradients_em}.

We notice - in agreement with the electrostatic findings of Section \ref{sec8} - that the inclusion of finite toroidal rotation leads to a mild peaking in the outer core regions ($\rho_{tor} = [0.3 - 0.6]$). 
Similarly as observed in Section \ref{sec8} in the electrostatic limit, the largest effect of a finite $E\times B$ is observed on the plasma density. As shown in Fig.~\ref{fig:pressure_em} and Fig.~\ref{fig:gradients_em}, it undergoes a large peaking throughout the radial domain $\rho_{tor} = [0.25 - 0.7]$.

\begin{figure}
\begin{center}
\includegraphics[scale=0.27]{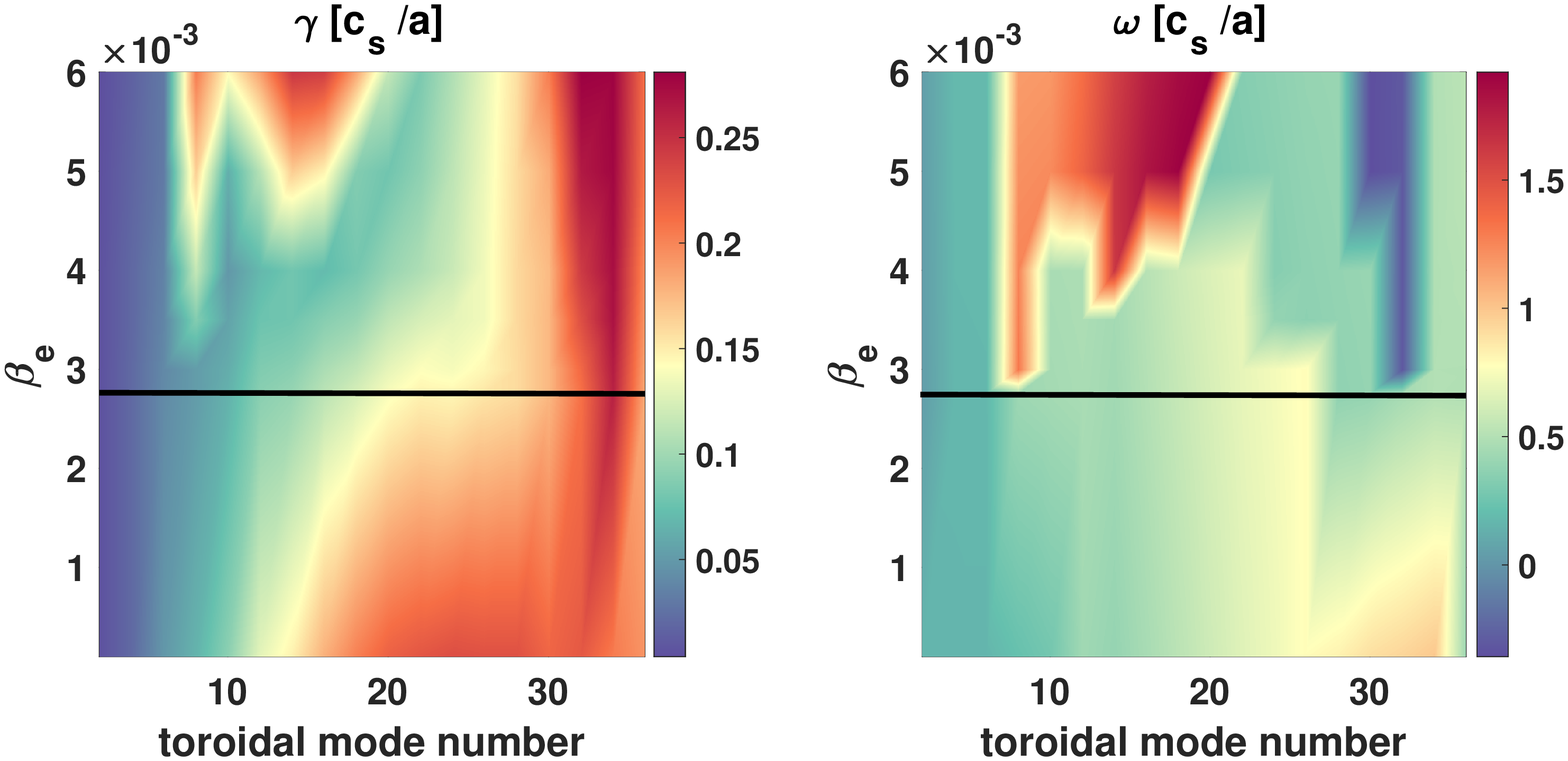}
\par\end{center}
\caption{Linear growth rates (a) and frequencies (b) computed by radially global linear GENE electromagnetic simulations on the steady-state GENE-Tango profiles for different toroidal mode-numbers and values of beta computed at the center of the GENE radial box. The value of the nominal plasma beta is marked by the horizontal black line.}
\label{fig:linear_EM}
\end{figure}

It is worth mentioning that the electromagnetic turbulence stabilization observed within this Section cannot be related to linearly unstable MHD modes. On the contrary, the steady-state profiles keep these modes sub-marginal. This is shown in Fig.~\ref{fig:linear_EM}, where the linear growth rates and frequencies of radially global linear simulations are shown for the final GENE-Tango profiles (see Fig.~\ref{fig:pressure_em}) for different toroidal mode-numbers and values of beta computed at the center of the GENE radial box. The value of the nominal plasma beta is marked in Fig.~\ref{fig:pressure_em}) by the horizontal black line.

%

\section{GENE-Tango speedups at ASDEX Upgrade and extrapolations to ITER}  \label{sec11}

Particular attention has been paid throughout this work to optimize the overall GENE run time during the GENE-Tango simulations for the ASDEX Upgrade discharge $\#31555$ at $t = 1.45$s. We found that the GENE-Tango convergence can be improved by applying different relaxation coefficients on the plasma profiles ($\alpha_p$) and turbulent fluxes ($\alpha_q$). More precisely, we used $\alpha_q = 0.4$ and $\alpha = [0.05 - 0.2]$. A larger value of $\alpha_q$ allows Tango to react more promptly to changes on the turbulent fluxes, thus adjusting the pressure profiles accordingly. On the other hand, the relaxation coefficient acting on the profiles had to be reduced to achieve faster and more stable iterations. This is due to the large sensitivity of the turbulent fluxes observed on even minor changes of the plasma pressure. The convergence speed depends strongly on the Tango extrapolation regions. Their location needs to be carefully adjusted to avoid nonphysical changes of the plasma profiles at the GENE buffer regions, which will inevitably affect the plasma pressure in the core. With these settings, the GENE run time for the turbulent calculation for the ASDEX Upgrade discharge $\#31555$ was $t =150 c_s / a$.

The overall run time of the GENE-Tango simulations with these optimal settings provides an estimate of the overall speedup achieved by properly exploiting the time-scale separation between micro- and macroscopic physics in the core of magnetic confinement devices. This can be done by comparing the time simulated by GENE with the energy confinement time for this ASDEX Upgrade discharge. In particular, a single GENE simulation letting the plasma profiles evolve due to the effect of physical sources (flux-driven) is expected to converge on timescales comparable to few confinement times. The energy confinement time is defined as the ratio between the plasma stored energy and the overall integral of the injected sources. This measures $\tau_E \approx 50$ms for the ASDEX Upgrade discharge $\#31555$ at $t = 1.45$s. To compare more easily the speedup achieved for each of the different cases studied in this paper, every GENE-Tango simulation was started from the reference TGLF-ASTRA profiles. We obtained the following speedups with GENE-Tango on 16 IBM POWER9 AC922 nodes each with 4 Nvidia Volta V100 GPUs on Marconi100.

\renewcommand{\labelenumii}{\Roman{enumii}}
\begin{enumerate}[(i)]

    \item Section \ref{sec5}: electromagnetic with collisions and external $E \times B$ shear - the GENE-Tango run time was $11.7$ms. Therefore, we achieved a speedup of $\tau_E / \tau_{{\rm GENE}} \approx 4.3$. The simulation lasted approximately 72h.
    
    \item Section \ref{sec6}: electrostatic with collisions - the GENE-Tango run time was $8.1$ms. Therefore, we achieved a speedup of $\tau_E / \tau_{{\rm GENE}} \approx 6.2$. The simulation lasted approximately 24h.
    
    \item Section \ref{sec7}: electrostatic with collisions and external $E \times B$ shear - the GENE-Tango run time was $6.1$ms. Therefore, we achieved a speedup of $\tau_E / \tau_{{\rm GENE}} \approx 8.2$. The simulation lasted approximately 24h.
    
    \item Section \ref{sec8}: electrostatic without collisions - the GENE-Tango run time was $12.7$ms. Therefore, we achieved a speedup of $\tau_E / \tau_{{\rm GENE}} \approx 4$. The simulation lasted approximately 24h.
    
    \item Section \ref{sec9}: electromagnetic with collisions - the GENE-Tango run time was $11.7$ms. Therefore, we achieved a speedup of $\tau_E / \tau_{{\rm GENE}} \approx 4.3$. The simulation lasted approximately 72h.

\end{enumerate}

These results reveal a significant speedup achieved when exploiting the multiple time-step approach of coupling the global version of the code GENE to the transport-solver Tango. At ASDEX Upgrade, we reached the steady-state solution roughly a factor of 4-8 faster than the expected run time of a single (flux-driven) GENE simulation.

These findings are particularly encouraging when performing extrapolations to larger devices, such as JET or ITER, since the time-step separation between turbulence and transport timescales is expected to scale as $(1/\rho_*)^2$ and thus the GENE-Tango speedup as $(1/\rho_*)^3$ considering a fixed radial resolution relative to the gyroradius.
\begin{figure}
\begin{center}
\includegraphics[scale=0.40]{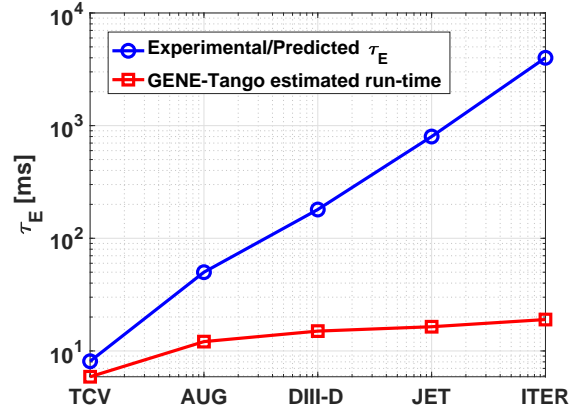}
\par\end{center}
\caption{Comparison of the confinement time measured at different tokamak devices, the H98$(y,2)$ ITER scaling \cite{ITER} and the estimated GENE run time in the GENE-Tango simulations. The estimated GENE-Tango run time has been computed assuming convergence is reached with 40 iterations each, corresponding to a GENE run time of $150 c_s /a$. The confinement time for the different devices and the reference electron temperature (at mid radius) required to compute the sound speed (and hence the overall GENE run time) are, respectively, taken from Ref.~\cite{Merlo_PPCF_2015} for TCV, from Ref.~\cite{Snyder_NF_2019} for DIII-D, from Ref.~\cite{Kim_NF_2020} for JET and from Ref.~\cite{Kinsey_NF_2011,Creely_JPP_2020} for ITER (H98$(y,2)$ ITER scaling). The ASDEX Upgrade GENE-Tango run-time has been set to $12$ms, which is the one obtained in the GENE-Tango simulation retaining the full physics of Section \ref{sec5}.}
\label{fig:speedup}
\end{figure}
In particular, assuming GENE-Tango would converge to the steady-state solution (turbulent fluxes match the volume integral of the injected sources) with comparable speed of the ASDEX Upgrade discharge $\# 31555$, we can compute a rough estimate for the overall GENE run time for different devices with GENE-Tango and compare this value to the confinement time. This is shown in Fig.~\ref{fig:speedup} by considering to reach convergence with 40 iterations each corresponding to a GENE run time of $150 c_s /a$. Fig.~\ref{fig:speedup} reveals a considerable increase in the GENE-Tango speedup (when compared to the confinement time, hence to a single flux-driven gyrokinetic simulation) as the size of the device increases. While this is negligible (only a factor of $\approx 1.5$) for TCV, it becomes significant for JET with an expected speedup of a factor of $\approx 15-18$. For the ITER standard scenario, this GENE-Tango convergence criterion leads to a speedup of roughly a factor of 200 compared to a single GENE (flux-driven) simulation for the same case and numerical setup. This is likely to be an underestimation and larger speedups are likely to be achieved. As discussed above, a time-dependent gyrokinetic simulations might require 2-3 confinement times to reach the steady-state solution, thus increasing the GENE-Tango speedup of ITER up to a factor of 600. This would make high-fidelity gyrokinetic simulations up to the transport time-scale possible with the current computing resources also for ITER-like devices for the first time.

\section{Conclusions} \label{sec12}

Global nonlinear simulations of different ASDEX Upgrade discharges have been performed with GENE-Tango up to the transport time-scale. We started our analyses by benchmarking GENE-Tango against previously published results obtained by coupling the flux-tube version of the GENE code to the transport-solver Trinity. The results refer to the ASDEX Upgrade shot $\# 13151$ at $t = 1.35$s. We found an excellent agreement between the ion and electron temperature profiles computed by GENE-Trinity and GENE-Tango with minor deviations in the plasma density. The analysis on the logarithmic gradients reveal that GENE-Trinity systematically under-predicts the temperature gradients with respect to GENE-Tango. This is likely to be caused by the local flux-tube approximation, which is known to over-predicts turbulent fluxes for medium-size devices such as ASDEX Upgrade \cite{Navarro_PoP_2016}, thus leading to reduced logarithmic gradients. The local and global turbulent fluxes are expected to converge as $1/\rho_* \rightarrow \infty$ \cite{McMillan_PRL_2010,goerler_PoP_2011}.

Afterwards, we analyzed the ASDEX Upgrade shot $\#31555$ at $t = 1.45$s with GENE-Tango. This plasma discharge shows a large peaking of the ion temperature profile, which is not captured by TGLF-ASTRA simulations. This is a known limitation for presently available reduced turbulence codes, which have been shown to strongly under-predict electromagnetic and supra-thermal particle effects on turbulence. These effects are typically found to reduce turbulent transport in gyrokinetic simulations consistently with the experimental signatures observed in different tokamak devices. We demonstrated that GENE-Tango can indeed recover the experimental measurements, correctly capturing the ion temperature peaking up to $\rho_{tor} \approx 0.2$. The correct description of the deep core regions ($\rho_{tor} < 0.2$) is likely to require the extension of GENE-Tango including supra-thermal particle physics. To assess the impact of $\beta$-stabilization, MHD modes, collisions and toroidal rotation on the evolution of the plasma profiles and $T_i$ peaking in the experiment, we performed GENE-Tango simulations retaining different physical effects in the GENE runs.

The main findings can be summarized as follows:

\renewcommand{\labelenumii}{\Roman{enumii}}
\begin{enumerate}[(i)]
    \item In the \textit{electrostatic limit}, GENE-Tango well reproduces the electron temperature profile, but shows a rather flat ion temperature and plasma density compared to ones computed with TGLF-ASTRA.  
    
    \item An excellent agreement between the \textit{electrostatic} GENE-Tango and TGLF-ASTRA profiles is obtained only when \textit{toroidal rotation} is retained in the GENE turbulent calculations. The inclusion of a finite $E\times B$ shear is found to stabilize turbulent transport in all different channels. However, the effect is mostly localized in the outer core regions $\rho_{tor} > 0.4$. Moreover, the largest effect of the toroidal rotation is observed on the plasma density, which undergoes a significant peaking for $\rho_{tor} > 0.4$. While these findings suggest that the TGLF model is essentially electrostatic, they hint that the significant on-axis ion temperature peaking observed in the experiment is not related to effects of the $E\times B$ shear, as also previously suggested in Ref.~\cite{Reisner_NF_2020}. In particular, both the GENE-Tango (with toroidal rotation) and the TGLF-ASTRA profiles do not exhibit this characteristic increase in the ion logarithmic temperature gradient.

    \item \textit{Collisions} play a major role in the evolution of the plasma density profile, partially affecting the temperatures (mainly electron). In particular, we found that collisions contribute to the particle flux by adding a positive contribution (namely a radially outward flux) that, together with the (positive-defined) diagonal diffusion, competes with the inward thermo-diffusion term. When collisions are neglected in the GENE-Tango simulations, the diagonal diffusion can no longer balance the thermo-diffusion contribution for the nominal profiles, resulting in negative particle flux. As a result, GENE-Tango strongly increases the amplitude of the logarithmic density gradient to enhance the diagonal diffusion until matching the volume integral of the injected particle flux (coming from both NBI and neutrals coming from the SOL). This process leads to a strong density peaking and a corresponding mild flattening of the electron temperature (due to the enhancement of TEM turbulence). These findings are well consistent with previous theoretical and experimental results, showing a favorable dependence of the density peaking with reduced collisionality \cite{Angioni_PRL_2003,Angioni_PoP_2003,Weisen_NF_2005,Greenwald_NF_2007,Angioni_NF_2007,Takenaga_NF_2008}.
    
    \item The peaking of the ion temperature profile observed in the experiment is recovered in the \textit{electromagnetic} GENE-Tango simulations with collisions. In particular, we observe in the initial GENE standalone simulation a significant turbulence suppression (respect to the electrostatic case) in all channels in the radial regions where the ion logarithmic temperature gradient increases in the experiment. This leads to a pronounced peaking in the ion profile computed by the GENE-Tango electromagnetic simulation. However, the on-axis temperature of the GENE-Tango ion temperature profile is still under-predicted with respect to the experimental measurements. This is likely to be caused by the absence of supra-thermal particles in the GENE simulations, which are well-known to stabilize ITG turbulence strongly when properly optimized. Fast ions will be included in the GENE-Tango workflow in the near future. Moreover, while the electron temperature is only mildly affected by electromagnetic fluctuations, we observe a flattening of the density profile. This is due to the enhancement of the outward flux of passing electrons in ITG turbulence as discussed in Ref.~\cite{Hein_PoP_2010}. We note that the steady-state plasma profiles do not excite any electromagnetic modes, which are submarginal, otherwise leading to a large increase in turbulent transport and possibly to a flattening of plasma profiles.

\end{enumerate}

\section*{Acknowledgement}

This work was supported by the U.S. Department of Energy under the Exascale Computing Project (17-SC-20-SC). This work has been carried out within the framework of the EUROfusion Consortium, funded by the European Union via the Euratom Research and Training Programme (Grant Agreement No 101052200 — EUROfusion). Views and opinions expressed are however those of the author(s) only and do not necessarily reflect those of the European Union or the European Commission. Neither the European Union nor the European Commission can be held responsible for them. This research used the resources of the Oak Ridge Leadership Computing Facility at the Oak Ridge National Laboratory, which is supported by the Office of Science of the U.S. Department of Energy under Contracts Nos. DE-AC05-00OR22725 and DE-AC02-05CH1123, Furthermore, numerical simulations were performed at the Marconi and Marconi100 Fusion supercomputers at CINECA, Italy.

\end{document}